\documentclass[amssymb,usenatbib,useAMS,usedcolumns]{mnras}
\usepackage{times} 
\usepackage{latexsym,amssymb,amsmath,amsfonts}
\usepackage{rotfloat} 
\usepackage{rotating} 
\usepackage{float}
\usepackage{graphicx}
\usepackage{natbib}
\usepackage{supertabular}
\usepackage{longtable}
\usepackage{url}
\usepackage{dcolumn}
\usepackage{placeins}
\usepackage{multirow} 
\usepackage{textcomp}
\usepackage{multicol}
\usepackage[caption=false]{subfig}
\begin{document}
\newcommand{\hi}{\mbox{H\,{\sc i}}}
\newcommand{\mgii}{\mbox{Mg\,{\sc ii}}}
\newcommand{\mgi}{\mbox{Mg\,{\sc i}}}
\newcommand{\feii}{\mbox{Fe\,{\sc ii}}}
\newcommand{\oi}{\mbox{O\,{\sc i}}}
\newcommand{\cii}{\mbox{C\,{\sc ii}}}
\newcommand{\ci}{\mbox{C\,{\sc i}}}
\newcommand{\sii}{\mbox{Si\,{\sc ii}}}
\newcommand{\znii}{\mbox{Zn~{\sc ii}}}
\newcommand{\mnii}{\mbox{Mn~{\sc ii}}}
\newcommand{\crii}{\mbox{Cr~{\sc ii}}}
\def\h2{$\rm H_2$}
\def\Nh2{$N$(H${_2}$)}
\def\chin{$\chi^2_{\nu}$}
\def\chiu{$\chi_{\rm UV}$}
\def\lya{\ensuremath{{\rm Ly}\alpha}}
\def\lymana{\ensuremath{{\rm Lyman}-\alpha}}
\def\kms{km\,s$^{-1}$}
\def\cms{cm$^{-2}$}
\def\cc{cm$^{-3}$}
\def\zabs{$z_{\rm abs}$}
\def\zem{$z_{\rm em}$}
\def\nhi{$N$($\hi$)}
\def\ln{log~$N$}
\def\nh{$n_{\rm H}$}
\def\ne{$n_{\rm e}$}
\def\21{21-cm}
\def\ts{$T_{\rm s}$}
\def\th{$T_{\rm 01}$}
\def\t0{$\tau_{\rm 0}$}
\def\ll{$\lambda\lambda$}
\def\l{$\lambda$}
\def\fc{$C_{\rm f}$}
\def\c21{$C_{21}^{\rm Fe~\textsc{ii}}$}
\def\mjb{mJy~beam$^{-1}$}
\def\taudv{$\int\tau dv$}
\def\ha{H\,$\alpha$}
\def\taudvl{$\int\tau dv^{3\sigma}_{10}$}
\def\taudv{$\int\tau dv$}
\def\wmg{$W_{\mgii}$}
\def\wfe{$W_{\feii}$}
\def\dgi{$\Delta (g-i)$}
\def\ebv{$E(B-V)$}
%
%
\title[Cold \hi\ gas in strong \feii\ absorbers]{
{Incidence of \hi\ \21\ absorption in strong \feii\ systems at $0.5<z<1.5$
}
\author[Dutta et al.]{R. Dutta$^1$\thanks{E-mail: rdutta@iucaa.in}, R. Srianand$^1$, N. Gupta$^1$, R. Joshi$^1$, P. Petitjean$^2$, P. Noterdaeme$^2$, J. Ge$^3$, 
\newauthor{J.-K. Krogager$^{2,4}$} \\
$^1$ Inter-University Centre for Astronomy and Astrophysics, Post Bag 4, Ganeshkhind, Pune 411007, India \\
$^2$ Institut d'Astrophysique de Paris, CNRS and UPMC Paris 6, UMR7095, 98bis boulevard Arago, 75014 Paris, France \\
$^3$ Astronomy Department, University of Florida, 211 Bryant Space Science Center, P.O. Box 112055, Gainesville, FL 32611, USA \\
$^4$ Dark Cosmology Centre, Niels Bohr Institute, University of Copenhagen, Juliane Maries Vej 30, 2100 Copenhagen \O{}, Denmark \\
}
}
\date{Accepted. Received; in original form }
\maketitle
\label{firstpage}
%
%
\begin {abstract}  
\par\noindent
We present the results from our search for \hi\ \21\ absorption in a sample of 16 strong \feii\ systems ($W_{\rm r}$(\mgii\ $\lambda$2796) $\ge$ 1.0 \AA\ 
and $W_{\rm r}$(\feii\ $\lambda$2600) or \wfe\ $\ge$ 1\,\AA) at $0.5<z<1.5$ using the Giant Metrewave Radio Telescope and the Green Bank Telescope. 
We report six new \hi\ \21\ absorption detections from our sample, which have increased the known number of detections in strong \mgii\ systems at 
this redshift range by $\sim$50\%. Combining our measurements with those in the literature, we find that the detection rate of \hi\ \21\ absorption 
increases with \wfe, being four times higher in systems with \wfe\ $\ge$ 1\,\AA\ compared to systems with \wfe\ $<$ 1\,\AA. The \nhi\ associated with 
the \hi\ \21\ absorbers would be $\ge 2 \times 10^{20}$\,\cms, assuming a spin temperature of $\sim$500~K (based on \hi\ \21\ absorption measurements 
of damped Lyman-$\alpha$ systems at this redshift range) and unit covering factor. We find that \hi\ \21\ absorption arises on an average in 
systems with stronger metal absorption. We also find that quasars with \hi\ \21\ absorption detected towards them have systematically higher \ebv\ 
values than those which do not. Further, by comparing the velocity widths of \hi\ \21\ absorption lines detected in absorption- and galaxy-selected 
samples, we find that they show an increasing trend (significant at 3.8$\sigma$) with redshift at $z<3.5$, which could imply that the absorption 
originates from more massive galaxy haloes at high-$z$. Increasing the number of \hi\ \21\ absorption detections at these redshifts is important 
to confirm various trends noted here with higher statistical significance.
\end {abstract}  
%
%
\begin{keywords} 
galaxies: quasar: absorption line $-$ galaxies: ISM    
\end{keywords}
%
%
\section{Introduction} 
\label{sec_introduction}  
The resonant absorption lines of \mgii\ detected towards background quasars have proved to be excellent probes of the gaseous haloes and circumgalactic medium 
(CGM) of $z\lesssim$ 2 galaxies in a luminosity-unbiased way \citep{lanzetta1987,sargent1988,petitjean1990,bergeron1991,steidel1992,srianand1994,steidel1995,nestor2005,prochter2006,quider2011,zhu2013}. 
While weak \mgii\ absorbers (i.e. absorbers having rest equivalent width of \mgii\ $\lambda 2796$, \wmg\ $<$ 1.0 \AA) could be tracing co-planar accreting or 
recycled gas \citep[e.g.][]{chen2010a,chen2010b,kacprzak2011a,kacprzak2011b,lovegrove2011}, strong \mgii\ absorbers (i.e. absorbers having \wmg\ $\ge$ 1.0 \AA) 
are believed to be tracing galactic winds or outflows \citep[e.g.][]{zibetti2007,bouche2007,weiner2009,lundgren2009,gauthier2009,noterdaeme2010b,rubin2010,menard2011,nestor2011}.
Strong \mgii\ absorbers at $z<$ 1.65 have been shown to trace gas with high neutral hydrogen column densities \citep[][hereafter R06]{rao2006}, like 
damped Lyman-$\alpha$ systems \citep[DLAs; \nhi\ $\ge$ $2\times10^{20}$\,\cms; see][for a review]{wolfe2005}. Further, the incidence of DLAs among \mgii\ 
absorbers is found to increase at high-$z$ \citep[2$<z<$6;][]{matejek2013}. \citet{murphy2007} have found a correlation between metallicity and \wmg\ in 
$z<2.6$ DLAs, similar to the velocity-metallicity relation seen in $z\ge1.8$ DLAs \citep{ledoux2006,prochaska2008a}. 

The strong \mgii\ systems in the \mgii\ absorber-galaxy catalog of \citet{nielsen2013} sample a wide range of galaxy impact parameters, i.e. over $\sim$10$-$200 kpc. 
Hence, such strong \mgii\ absorbers can trace gas in a wide variety of environments like star-forming discs, CGM, galactic winds and outflows. It is likely that 
\wmg\ is dominated by high column density when the sightlines probe the galactic discs and by velocity spread of the absorbing gas in the other scenarios. If one 
is interested in studying the cold dense gas around galaxies, other parameters like equivalent width ratios of metal lines are required to select sightlines that 
probe low impact parameters. R06 have demonstrated that equivalent width ratios of \mgii, \mgi\ and \feii\ absorption can be used to pre-select DLAs more successfully 
than by just using \wmg. They detect DLAs with a success rate of $\sim$42\% by selecting \mgii\ absorbers with $W_r(\mgii\ \lambda 2796)/W_r(\feii\ \lambda 2600) < 2$ 
and $W_r(\mgi\ \lambda 2852) > 0.1$~\AA. 

Further insights into the origin and physical conditions prevailing in the strong \mgii\ systems and DLAs can be obtained by studying their associated 
\hi\ \21\ absorption. \hi\ \21\ absorption is an excellent tracer of the cold neutral medium (CNM; $T\sim$ few 100 K) of galaxies \citep{kulkarni1988},
and has been observed to arise at small impact parameters ($<$ 30 kpc) from $z<$ 0.4 galaxies \citep{dutta2016b}. There have been numerous searches for 
\hi\ \21\ absorption in \mgii\ systems and DLAs \citep[e.g.][]{briggs1983,lane2000,kanekar2003,curran2005,gupta2009,kanekar2009a,curran2010a,srianand2012,gupta2012,kanekar2014a}. 
\citet[][hereafter G09 and G12, respectively]{gupta2009,gupta2012} have shown that the \hi\ \21\ detection rate in strong \mgii\ systems can be enhanced 
with appropriate equivalent width ratio cuts of \mgii, \feii\ and \mgi. 

In addition, searches of \hi\ \21\ absorption in samples of \mgii\ systems and DLAs can be used to trace the redshift evolution of the CNM fraction in galaxies. 
Spin temperature (\ts) measurements and upper limits, derived using \hi\ \21\ optical depth and \nhi\ measured from DLAs, suggest that most of the gas along 
these sightlines traces the diffuse warm neutral medium (WNM; $T\sim$ 10$^4$ K), and a small fraction of the total \nhi\ along these sightlines is 
associated with the CNM \citep{srianand2012,kanekar2014a}. This is also supported by observations of typically low overall molecular fractions of \h2\ 
in $z>1.8$ DLAs \citep{petitjean2000,ledoux2003,srianand2005,noterdaeme2008}, and physical conditions inferred in DLAs using C\,{\sc ii}* and Si\,{\sc ii}* 
fine structure lines \citep{neeleman2015}. The above studies have indicated that the CNM covering factor in DLAs and \mgii\ systems may be small ($\sim$10-20\%). 
On the other hand, \citet{heiles2003}, using \hi\ \21\ emission and absorption measurements, have found that $\sim$40\% of the \hi\ gas is in the CNM in the 
Milky Way. Since the volume filling factor and the physical conditions of the CNM depend on feedback processes related to the {\it in situ} star formation 
\citep{mckee1977,wolfire1995,deavillez2004,gent2013,gatto2015}, understanding the redshift evolution of \hi\ \21\ absorbers can shed light on the star 
formation history of galaxies as well. 

In this work we wish to explore the connection between \mgii\ systems, DLAs and cold \hi\ gas by searching for \hi\ \21\ absorption in a sample of strong 
\mgii\ systems, that satisfy additional constraints to increase the efficiency of detecting cold gas. Systems showing Si\,{\sc ii} or C\,{\sc i} absorption 
would have been excellent targets to search for cold gas, since strong Si\,{\sc ii} absorption is likely to select high metallicity gas \citep{prochaska2008a,jorgenson2013}, 
whereas C\,{\sc i} absorption is likely to probe dense molecular regions \citep{noterdaeme2011,noterdaeme2016}. However, the strong transitions of Si\,{\sc ii} 
$\lambda$1526 and C\,{\sc i} $\lambda$1656 are not covered simultaneously with the \mgii\ doublet lines of the same system for a large redshift range. On 
the other hand, similar to the \mgii\ doublet, strong \feii\ lines, like the $\lambda$2600 \AA\ transition, can be observed with ground-based telescopes at 
0.4$<z<$2.5. 

Strong \feii\ systems can arise from either very high metallicity sub-DLAs or high \nhi\ DLAs \citep{srianand1996}. The \hi\ \21\ optical depth is directly 
proportional to \nhi\ and inversely to \ts. There are indications for a weak anti-correlation between \ts\ and the gas phase metallicity \citep[see][and references therein]{kanekar2014a}. 
Further, strong absorption from \mgii\ (\wmg\ $\ge$ 1 \AA) and \feii\ (rest equivalent width of \feii\ $\lambda$2600, \wfe\ $\ge$ 1 \AA) are usually detected  
whenever galaxy nebular emission lines are directly detected in the Sloan Digital Sky Server \citep[SDSS;][]{york2000} quasar spectra, indicating that impact 
parameters are $\lesssim$10 kpc \citep[][Joshi et al. 2016, in preparation]{noterdaeme2010b}. We also notice that $\sim$77\% of the \mgii\ systems showing 
signatures of 2175 \AA\ bump studied by \citet{jiang2011} have \wfe\ $\ge$ 1 \AA. Hence, systems with strong \feii\ absorption could provide ideal targets 
for detecting high metallicity and/or high \nhi\ cold gas.

This paper is structured as follows. In Section \ref{sec_sample}, we describe the motivation behind our sample selection and give details of our sample. 
In Section \ref{sec_observations}, we describe our radio observations using the Giant Metrewave Radio Telescope (GMRT) and the Green Bank Telescope (GBT). 
The results from our search for \hi\ \21\ absorption are presented in Section \ref{sec_parameters}. The detection rate of \hi\ \21\ absorption in strong 
\feii\ systems is discussed in Section \ref{sec_coveringfactor}. The connection of \hi\ \21\ absorption with metal line properties and dust content are 
discussed in Sections \ref{sec_metals} and \ref{sec_dust}, respectively. Finally, we study the velocity width of the \hi\ \21\ absorption lines in Section \ref{sec_velocity}. 
We summarize our results in Section \ref{sec_summary}. Throughout this work we have adopted a flat $\Lambda$-cold dark matter cosmology with $H_{\rm 0}$ = 70\,\kms~Mpc$^{-1}$ and $\Omega_M$ = 0.30.
%
%
\section{Sample Selection}  
\label{sec_sample}  
\begin{figure*}
\includegraphics[height=0.35\textheight, angle=90]{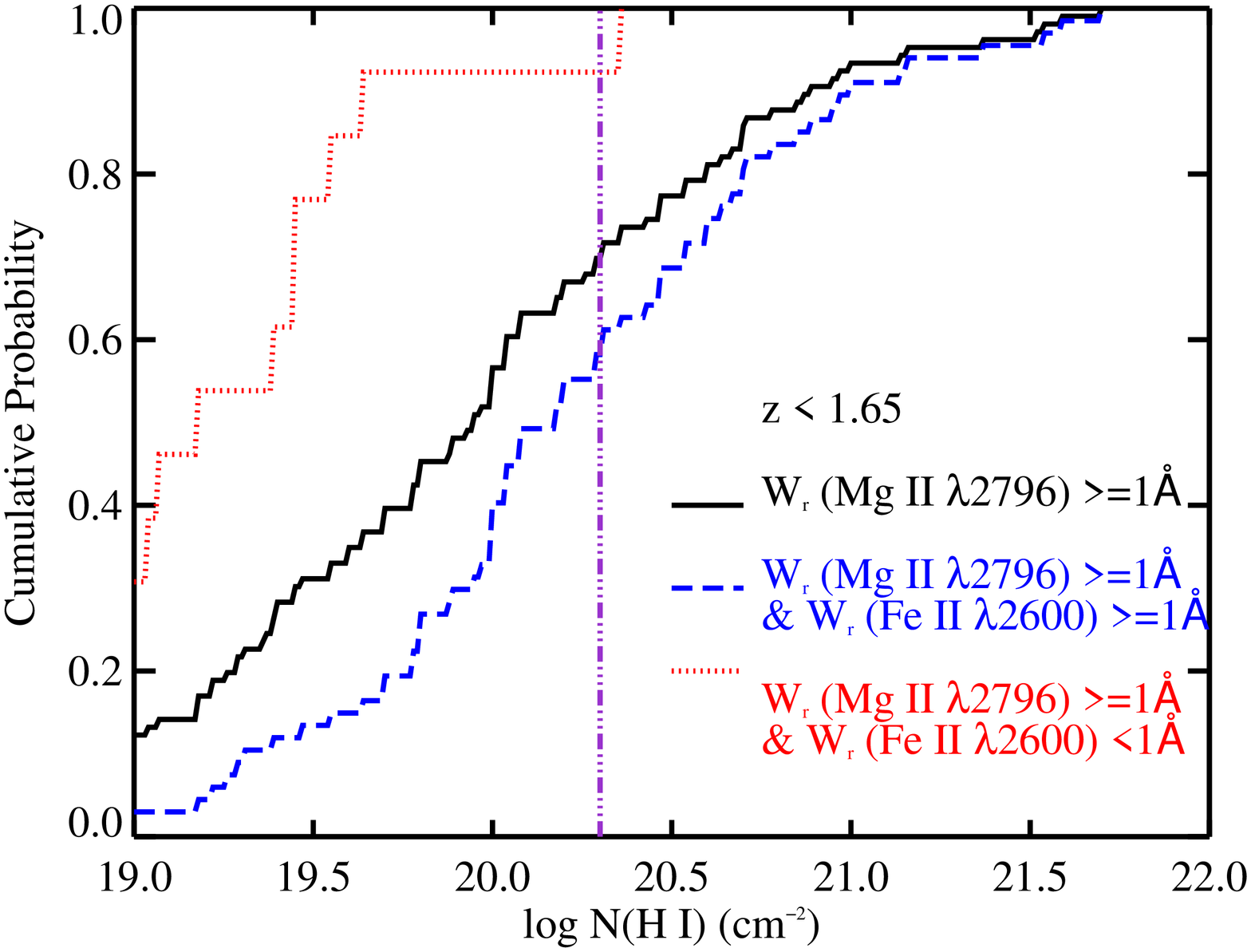} 
\includegraphics[height=0.35\textheight, angle=90]{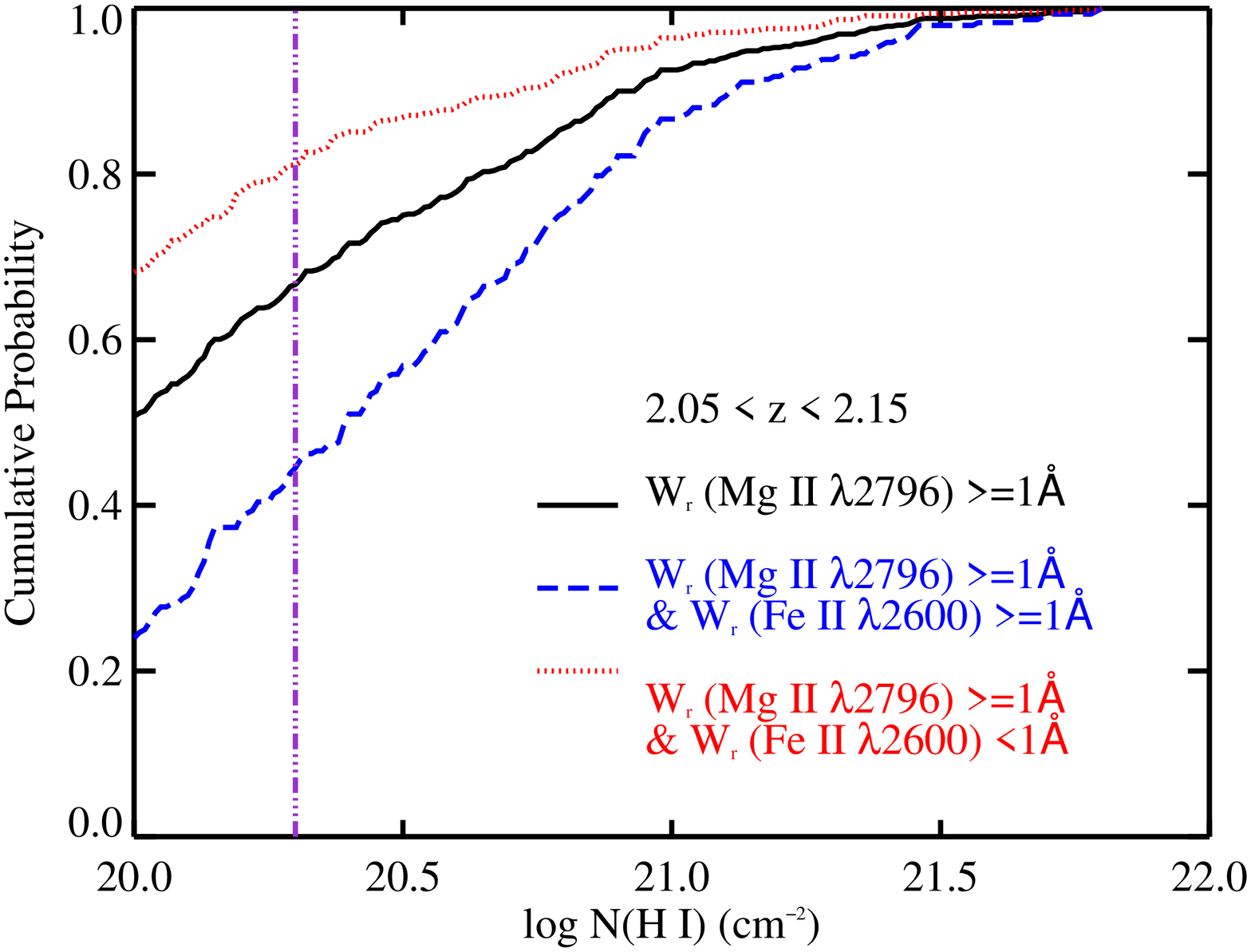}
\caption{The cumulative distribution of log~\nhi\ in systems with \wmg\ $\ge1$ \AA\ (solid), those with both \wmg\ $\ge1$ \AA\ and \wfe\ $\ge1$ \AA\ (dashed), 
and those with \wmg\ $\ge1$ \AA\ and \wfe\ $<1$ \AA\ (dotted).
The left panel shows the distribution for systems at $z<$ 1.65 with \nhi\ measurements from R06 using HST.
The right panel shows the distribution for systems at 2.05$<z<$2.15 from \citet{zhu2013} with \nhi\ measurements from \citet{noterdaeme2012} using SDSS.
The vertical line in each panel marks \nhi\ = $2 \times 10^{20}$\,\cms.}
\label{fig:cumdist}
\end{figure*}

While \mgii\ absorption lines at 0.3 $<z<$ 2.3 can be observed using ground-based optical telescopes, the atmospheric cutoff of light below 3000 \AA\ 
restricts ground-based observations of \hi\ \lymana\ lines at $z<$ 1.5. However, ultraviolet (UV) spectroscopic observations with the Hubble Space 
Telescope (HST) make it possible to detect $z<$ 1.65 DLAs and sub-DLAs \citep[R06,][]{rao2011,meiring2011,battisti2012,turnshek2015,neeleman2016}. 
R06 have obtained UV observations of \lymana\ absorption in 197 \mgii\ absorbers with \wmg\ $\ge$ 0.3 \AA\ at $z<$ 1.65, using the HST. Using their 
table 1, we plot the cumulative distribution of \nhi\ in absorbers with \wmg\ $\ge1$ \AA\ in left panel of Fig.~\ref{fig:cumdist}. We also show 
the same for systems with both \wmg\ $\ge1$ \AA\ and \wfe\ $\ge1$ \AA, and for those with \wmg\ $\ge1$ \AA\ but \wfe\ $<1$ \AA \footnote{For the 
cumulative distribution of systems with \wmg\ $\ge1$ \AA\ and \wfe\ $<1$ \AA, we use the subsamples 1 and 2 of R06, which were selected purely
based on \wmg\ without any constraint on \wfe\ (see R06 for details), to avoid biases.}. It can be seen that 32\% of the \mgii\ absorbers with 
\wmg\ $\ge$ 1.0 \AA\ have \nhi\ $\ge 2 \times 10^{20}$\,\cms. This fraction becomes 45\% when the strong \mgii\ absorbers also have \wfe\ $\ge$ 
1.0 \AA, i.e. strong \feii\ absorbers have a higher (by a factor of $\sim$1.4) probability of being DLAs. For comparison only 8\% of the absorbers 
with \wmg\ $\ge$ 1.0 \AA\ and \wfe\ $<$ 1.0 \AA\ are DLAs. We note that the success rate of a \wfe\ $\ge$ 1.0 \AA\ selection in detecting DLAs is 
consistent with that of a \wmg$/$\wfe\ $<$ 2 selection suggested by R06 (see Section~\ref{sec_introduction}).

Next, from the SDSS \mgii\ absorber catalog of \citet{zhu2013}, we identified 741 \mgii\ absorbers at $2.05<z<2.15$ with \wmg\ $\ge$ 1.0 \AA, which 
have been searched for \lymana\ absorption by \citet{noterdaeme2012}. In the right panel of Fig.~\ref{fig:cumdist} we show the cumulative distributions 
of \nhi\ in these systems, for the same three equivalent width cuts of \mgii\ and \feii\ as described above. Note that \citet{noterdaeme2012} 
provides \nhi\ measurements only for systems having \nhi\ $\ge$ $10^{20}$\,\cms. Hence, we show the cumulative distributions starting at \nhi\ 
= $10^{20}$\,\cms. Similar to above, we find that an additional cut of \wfe\ $\ge$ 1 \AA\ increases the probability of selecting DLAs by a factor 
of $\sim$1.7 compared to just a \wmg\ $\ge$ 1 \AA\ cut. It can be seen that 57\% of the strong \feii\ absorbers are DLAs, compared to 34\% of the 
absorbers with just \wmg\ $\ge$ 1.0 \AA. Further, only 19\% of the absorbers with \wmg\ $\ge$ 1.0 \AA\ and \wfe\ $<$ 1.0 \AA\ are DLAs. 

From the above discussions it seems that a selection technique based on strong \feii\ absorption has higher (by a factor of $\sim$1.4$-$1.7) probability of detecting DLAs than that based 
solely on \mgii\ absorption strength. Though, as can be seen from Fig.~\ref{fig:cumdist}, for a limiting rest equivalent width of \feii, the DLA selection efficiency appears to decrease 
slightly with decreasing redshift. Similarly, \citet{matejek2013}, using infrared spectra of high-$z$ quasars, have found that the fraction of DLAs in \mgii\ systems with \wmg\ $\ge$ 0.3 
\AA\ increases from $\sim$17\% at $z<$ 2 to $\sim$41\% at $z\ge$ 2. The metallicity in DLAs is known to decline with increasing redshift \citep[e.g.][]{rafelski2014}. Hence, for a given
equivalent width of \mgii\ or \feii, the same absorber is likely to arise from a higher \nhi\ system at high-$z$ compared to at low-$z$. This could be a possible explanation for a larger 
fraction of the strong \feii$/$\mgii\ absorbers being DLAs at higher redshifts.

We set out to quantify the \hi\ \21\ absorption detection efficiency in such a strong \feii\ selected sample. Our sample of strong \feii\ absorbers is drawn from the \mgii\ catalog of 
\citet{zhu2013} constructed using the SDSS Data Release 12 \citep[DR12;][]{alam2015}. We selected all the absorbers with \wmg\ $\ge$ 1.0 and \wfe\ $\ge$ 1.0, over the redshift ranges 
0.5$< z<$1.0 and 1.1$< z<$1.5. Next, by cross-correlating the positions of the background quasars with radio sources within $\sim$1$''$ radius in the Faint Images of the Radio Sky at 
Twenty-Centimeters \citep[FIRST;][]{white1997} survey, we identified 17 and 18 strong \feii\ systems at 0.5$< z<$1.0 and 1.1$< z<$1.5, respectively, that are in front of quasars with peak 
flux density at 1.4 GHz greater than 100 mJy. The redshift ranges of 0.5$< z<$1.0 and 1.1$< z<$1.5 were chosen to enable observations with the GBT prime-focus PF1 800 MHz receiver 
and the GMRT 610 MHz receiver, respectively (see Section~\ref{sec_observations}). The flux density cut of 100 mJy was used in order to achieve 3$\sigma$ integrated \hi\ \21\ optical 
depth sensitivity (for a line width of 10 \kms) $\lesssim$0.3\,\kms, which is sensitive to detect 100 K gas with \nhi\ $\le$ 5$\times$10$^{19}$\,\cms. 

We visually inspected the optical spectra and radio images of the quasars to ensure that there are no false identifications. Among the 35 strong \feii\ absorbers, 11 are part of \mgii\ 
samples that have been searched for \hi\ \21\ absorption in the literature $-$ 4 in G09, 3 in G12, 3 in \citet{kanekar2009a} (hereafter K09) and 1 in \citet{lane2000} (hereafter L00).
From the remaining 24 systems, we selected 15 strong \feii\ absorbers, whose redshifted \hi\ \21\ frequencies fall in the radio frequency interference (RFI)-free observing bands of GBT and 
GMRT. In addition, we included the \zabs\ = 1.27384 \mgii\ absorber towards J091927.61$+$014603.0. Note that this system is not in the catalog of \citet{zhu2013} and the peak flux density
of the radio source from FIRST is less than 100 mJy (i.e. 59 mJy). But this system satisfies our criterion of \wmg\ $\ge$ 1.0 and \wfe\ $\ge$ 1.0, and has been identified as a 2175 \AA\ 
absorber candidate by \citet{jiang2011}. Hence, our sample consists of 16 absorbers, of which seven (at 0.5$<z<$1.0) were observed with GBT, and nine (at 1.1$<z<$1.5) were observed with GMRT. 
The details of these 16 absorbers are presented in Table~\ref{tab:sample}. The median values of various metal line parameters for our observed sample are: \wmg\ = 2.23 \AA; \wfe\ = 1.55 \AA; 
\mgii\ doublet ratio, DR = 1.14; $W_{\rm r}$(\mgi\ $\lambda$2852)$/$\wmg, R1 = 0.27; \wmg$/$\wfe, R2 = 1.58.

\begin{table*} 
\caption{Sample of strong \feii\ systems}
\centering
\begin{tabular}{cccccccc}
\hline
Quasar & SDSS  & \zem\ & \zabs\ & $W_{\rm r}$(\mgii\ $\lambda$2796) & $W_{\rm r}$(\mgii\ $\lambda$2803) & $W_{\rm r}$(\mgi\ $\lambda$2852) & $W_{\rm r}$(\feii\ $\lambda$2600) \\
name   & name  &       &        & (\AA)                             & (\AA)                             & (\AA)                            & (\AA)                             \\
(1)    & (2)   & (3)   & (4)    & (5)                               & (6)                               & (7)                              & (8)                               \\
\hline
J0919$+$0146 & J091927.61$+$014603.0 & 1.286 & 1.2738 & 2.70 $\pm$ 0.15 & 1.99 $\pm$ 0.15 & 0.27 $\pm$ 0.13 & 2.24 $\pm$ 0.28 \\
J0921$+$6215 & J092136.22$+$621552.5 & 1.447 & 1.1035 & 4.11 $\pm$ 0.10 & 3.62 $\pm$ 0.13 & 0.98 $\pm$ 0.10 & 2.86 $\pm$ 0.09 \\ 
J0952$+$5048 & J095227.30$+$504850.6 & 1.091 & 0.9985 & 2.10 $\pm$ 0.04 & 1.94 $\pm$ 0.04 & 0.40 $\pm$ 0.05 & 1.30 $\pm$ 0.04 \\ 
J1111$+$3252 & J111131.77$+$325255.8 & 2.362 & 1.4753 & 2.23 $\pm$ 0.13 & 2.35 $\pm$ 0.14 & 0.96 $\pm$ 0.10 & 1.55 $\pm$ 0.12 \\
J1111$+$4507 & J111114.81$+$450735.0 & 1.486 & 0.6865 & 1.83 $\pm$ 0.14 & 1.93 $\pm$ 0.14 & 0.83 $\pm$ 0.15 & 1.73 $\pm$ 0.16 \\
J1201$+$1114 & J120140.25$+$111447.6 & 2.296 & 1.4846 & 2.66 $\pm$ 0.14 & 2.40 $\pm$ 0.13 & 0.31 $\pm$ 0.10 & 1.57 $\pm$ 0.12 \\
J1241$+$6020 & J124129.57$+$602041.3 & 2.066 & 1.2379 & 4.73 $\pm$ 0.06 & 4.48 $\pm$ 0.06 & 1.34 $\pm$ 0.07 & 3.38 $\pm$ 0.08 \\
J1245$+$2232 & J124557.77$+$223205.3 & 1.478 & 0.5808 & 1.64 $\pm$ 0.05 & 1.40 $\pm$ 0.05 & 0.16 $\pm$ 0.04 & 1.14 $\pm$ 0.05 \\
J1255$+$1817 & J125531.75$+$181750.9 & 1.371 & 0.7580 & 2.20 $\pm$ 0.18 & 1.88 $\pm$ 0.18 & 0.71 $\pm$ 0.17 & 1.23 $\pm$ 0.18 \\
J1327$+$4326 & J132720.97$+$432627.9 & 2.086 & 0.9542 & 2.42 $\pm$ 0.12 & 2.09 $\pm$ 0.10 & 0.47 $\pm$ 0.16 & 1.53 $\pm$ 0.11 \\
J1342$+$5110 & J134224.31$+$511012.4 & 2.598 & 1.4880 & 3.18 $\pm$ 0.07 & 2.92 $\pm$ 0.07 & 1.06 $\pm$ 0.07 & 1.74 $\pm$ 0.06 \\
J1351$+$0830 & J135116.91$+$083039.8 & 1.442 & 1.4270 & 2.59 $\pm$ 0.05 & 2.09 $\pm$ 0.05 & 0.85 $\pm$ 0.08 & 1.43 $\pm$ 0.07 \\
J1504$+$2854 & J150426.69$+$285430.5 & 2.285 & 1.2208 & 1.87 $\pm$ 0.07 & 1.58 $\pm$ 0.07 & 0.50 $\pm$ 0.08 & 1.06 $\pm$ 0.07 \\
J1510$+$1640 & J151038.63$+$164010.2 & 1.827 & 1.2139 & 2.22 $\pm$ 0.09 & 1.70 $\pm$ 0.08 & 0.34 $\pm$ 0.09 & 1.29 $\pm$ 0.09 \\
J2219$+$0229 & J221930.79$+$022945.4 & 2.218 & 0.9804 & 2.19 $\pm$ 0.12 & 2.08 $\pm$ 0.12 & 0.64 $\pm$ 0.12 & 1.78 $\pm$ 0.13 \\
J2330$+$1100 & J233040.84$+$110018.6 & 1.502 & 0.9488 & 1.55 $\pm$ 0.05 & 1.32 $\pm$ 0.05 & 0.40 $\pm$ 0.06 & 1.07 $\pm$ 0.05 \\
\hline
\end{tabular}
\label{tab:sample}
\begin{flushleft} {\it Notes.}
Column 1: quasar name used in this work. Column 2: SDSS name (J2000) of the quasar. Column 3: redshift of quasar. Column 4: redshift of intervening \feii\ system. 
Columns 5, 6, 7 and 8: rest frame equivalent widths (in \AA) of the \mgii\ $\lambda$2796, \mgii\ $\lambda$2803, \mgi\ $\lambda$2852 and \feii\ $\lambda$2600 absorption 
lines, respectively, taken from \citet{zhu2013} (except in case of J0919$+$0146, where we measure these ourselves). \\
\end{flushleft}
\end{table*}
The 1.4 GHz peak flux densities obtained by the FIRST survey of the radio sources in our sample are listed in Table~\ref{tab:radiosource}. All the sources except one are compact 
\citep[deconvolved sizes $\le$2$''$;][]{white1997} in the FIRST images (resolution $\sim$5$''$). The radio source J1510$+$1640 is resolved in the FIRST image, with a deconvolved size of 
2.9$''\times$1.3$''$, and with the peak emission accounting for 86\% of the total emission. Low-frequency ($\le$1.4 GHz) sub-arcsecond-scale images are not available for any of the sources 
in our sample. However, 2.3 GHz Very Long Baseline Array (VLBA) images are available for 8 of the sources from the VLBA Calibrator Survey (VCS)\footnote{http://www.vlba.nrao.edu/astro/calib/}. 
We assume that the covering factor (\fc) of the absorbing gas is equal to the core fraction, i.e. we assume that the gas covers only the core component seen in VLBA images. Hence, 
we estimate \fc\ from the ratio of the peak flux density in the VCS image to the total arcsecond-scale flux density (see Table~\ref{tab:radiosource}). The total arcsecond-scale 
flux densities at 2.3 GHz have been estimated by interpolating flux densities available from the NASA extragalactic database \footnote{https://ned.ipac.caltech.edu/} \citep{condon1998,gregory1991}. 
\begin{table} 
\caption{Details of the radio sources}
\centering
\begin{tabular}{cccc}
\hline
Quasar & 1.4 GHz Peak Flux & Morph. & \fc\ \\
       & Density (\mjb)    &        &      \\
(1)    & (2)               & (3)    & (4)  \\
\hline
J0919$+$0146 &   59 & C & ---  \\  
J0921$+$6215 & 1213 & C & 0.60 \\
J0952$+$5048 &  105 & C & 0.71 \\
J1111$+$3252 &  107 & C & ---  \\
J1111$+$4507 &  239 & C & ---  \\
J1201$+$1114 &  161 & C & ---  \\
J1241$+$6020 &  433 & C & 0.88 \\
J1245$+$2232 &  131 & C & ---  \\
J1255$+$1817 &  366 & C & 0.36 \\
J1327$+$4326 &  589 & C & 0.91 \\
J1342$+$5110 &  144 & C & ---  \\
J1351$+$0830 &  335 & C & 0.92 \\
J1504$+$2854 &  549 & C & 0.60 \\
J1510$+$1640 &  138 & R & ---  \\ 
J2219$+$0229 &  183 & C & ---  \\
J2330$+$1100 & 1174 & C & 0.59 \\
\hline
\end{tabular}
\label{tab:radiosource}
\begin{flushleft} {\it Notes.}
Column 1: quasar name. Column 2: FIRST 1.4 GHz peak flux density in \mjb. 
Column 3: morphology from FIRST image ($\sim$5$''$ resolution), where C is for compact and R is for resolved.
Column 4: covering factor (assumed to be the core fraction) determined from 2.3 GHz VLBA images (see Section~\ref{sec_sample} for details). \\
\end{flushleft}
\end{table}
%
%
\section{Radio Observations}  
\label{sec_observations}  
The seven systems at 0.5$< z<$1.0 were observed with the prime-focus PF1 800 MHz receiver at GBT (Proposal ID: 16A$-$141). 
We used VEGAS as the backend with a bandwidth of 11.72 MHz split into 32768 channels (spectral resolution of $\sim$0.1\,\kms\ 
per channel, velocity coverage $\sim$4000\,\kms). The observations were carried out in standard position switching mode, with 
$\sim$2 min spent on-source and same amount of time spent at the off-source position. A dump time of 2 s was used. The data were 
acquired in two linear polarization products, XX and YY. The nine systems at 1.1$< z<$1.5 were observed with the 610 MHz receiver 
at GMRT, using the 2 MHz baseband bandwidth split into 512 channels (spectral resolution $\sim$2\,\kms\ per channel, velocity coverage 
$\sim$1000\,\kms). Data were acquired in two polarization products, LL and RR. Standard calibrators were regularly observed during 
the observations for flux density, bandpass, and phase calibrations. In both the GBT and the GMRT observations, the pointing centre 
was at the quasar coordinates and the band was centred at the redshifted \hi\ \21\ frequency. 

The GMRT data were reduced using the National Radio Astronomy Observatory (NRAO) Astronomical Image Processing System ({\sc aips}) 
following standard procedures as described in \citet{gupta2010}. The GBT data were reduced using a pipeline developed using NRAO's 
{\sc gbtidl} package as described in G12. The details of the radio observations of the sources are given in Table~\ref{tab:obslog}.

All the nine radio sources observed with GMRT, except J1510$+$1640, are compact in our GMRT images. Note that the sources, J1111$+$3252 
(deconvolved size $\sim$2.2$''\times$0.2$''$) and J1351$+$0830 (deconvolved size $\sim$2.0$''\times$0.9$''$), are considered as 
compact since $>$90\% of the total emission is contained in a single Gaussian component. In case of J1510$+$1640, the synthesized 
beam of our GMRT image is $\sim$6.2$''\times$3.9$''$, and the radio emission can be represented by a single Gaussian component 
with deconvolved size $\sim$2.9$''\times$0.5$''$, which contains $\sim$81\% of the total emission, similar to the FIRST image. 
Note that in all the GMRT data cubes, the \hi\ \21\ absorption spectra were extracted at the location of the continuum peak flux density.

In case of seven radio sources observed with GBT, we obtained the flux density of five of them from the GBT spectrum. For two sources, 
J1255$+$1817 and J2219$+$0229, the flux scale could not be calibrated properly in our GBT observations. For these sources, the flux densities 
at the redshifted \hi\ \21\ frequency have been estimated by interpolating between the flux densities from the FIRST and the TEXAS \citep{douglas1996} 
surveys at 1.4 GHz and 365 MHz, respectively. The flux densities at the redshifted \hi\ \21\ frequency of all the radio sources observed by 
us are given in Table~\ref{tab:radiopara}.
\begin{table} 
\caption{Radio observation log}
\centering 
\begin{tabular}{cccc}
\hline
Quasar & Date & Time & $\delta v$ \\
       &      & (h)  & (\kms)     \\
(1)    & (2)  & (3)  & (4)        \\
\hline
\multicolumn{4}{c}{GBT} \\
\hline
J0952$+$5048 & 18 February 2016 & 1.7 & 0.2 \\ 
             & 13 March 2016    & 0.7 & 0.2 \\
             & 21 May 2016      & 0.7 & 0.2 \\
J1111$+$4507 & 24 February 2016 & 1.0 & 0.1 \\
             & 11 March 2016    & 0.3 & 0.1 \\
J1245$+$2232 & 18 February 2016 & 0.8 & 0.1 \\
J1255$+$1817 & 18 February 2016 & 0.6 & 0.1 \\
J1327$+$4326 & 24 February 2016 & 0.5 & 0.1 \\
             & 11 March 2016    & 0.2 & 0.1 \\ 
             & 13 March 2016    & 0.6 & 0.1 \\
J2219$+$0229 & 23 February 2016 & 0.5 & 0.1 \\
             & 11 March 2016    & 0.4 & 0.1 \\
J2330$+$1100 & 23 February 2016 & 0.3 & 0.1 \\
\hline
\multicolumn{4}{c}{GMRT} \\
\hline
J0919$+$0146 & 24 November 2015 & 6.4 & 2.0 \\
             & 12 June 2016     & 4.2 & 2.0 \\
J0921$+$6215 & 25 November 2015 & 0.8 & 1.8 \\ 
J1111$+$3252 & 31 October 2015  & 5.5 & 2.1 \\
J1201$+$1114 & 14 December 2015 & 5.1 & 2.1 \\
J1241$+$6020 & 22 November 2015 & 3.2 & 1.9 \\
J1342$+$5110 & 21 November 2015 & 5.2 & 2.1 \\
             & 26 April 2016    & 5.4 & 2.1 \\
J1351$+$0830 & 22 January 2016  & 3.2 & 2.1 \\
J1504$+$2854 & 25 November 2015 & 3.2 & 1.9 \\
J1510$+$1640 & 4 January 2016   & 2.9 & 1.9 \\
             & 5 January 2016   & 3.2 & 1.9 \\
\hline
\end{tabular}
\label{tab:obslog} 
\begin{flushleft} {\it Notes.}
Column 1: quasar name. Column 2: date of observation. Column 3: time on source in h. Column 4: channel width in \kms. \\
\end{flushleft}
\end{table}
%
%
\section{Parameters derived from \hi\ \21\ absorption spectra}
\label{sec_parameters}
The results from our search for \hi\ \21\ absorption in the 16 strong \feii\ absorbers are summarized in Table~\ref{tab:radiopara}. 
We have detected \hi\ \21\ absorption in 6 out of these 16 strong \feii\ systems. We provide the standard deviation in the optical 
depth at $\sim$2\,\kms\ spectral resolution ($\sigma_\tau$), and 3$\sigma$ upper limit on the integrated optical depth from spectra 
smoothed to 10\,\kms\ (\taudvl). In case of \hi\ \21\ absorption detections, we provide the peak optical depth ($\tau_{\rm p}$) and 
the total integrated optical depth (\taudv). From \citet{kanekar2014a} we see that the mean and median \ts\ in DLAs over 0.5$<z<$1.5 
are 670 K and 460 K, respectively. We estimate \nhi\ from \taudv\ of the detections, or 3$\sigma$ upper limit to it from \taudvl\ in 
case of non-detections, assuming a typical \ts\ of 500 K and \fc\ = 1. The \nhi\ associated with the \hi\ \21\ absorption will be 
$\ge 2 \times 10^{20}$ \cms\ under these assumptions, i.e. these systems will satisfy the definition of being DLAs. The optical depth 
sensitivity in 8 out of the 10 cases of \hi\ \21\ non-detection rules out the system being DLA if \ts\ = 500 K and \fc\ = 1. 
Table~\ref{tab:radiopara} also gives the velocity width which contains 90\% of the total optical depth ($v_{\rm 90}$), and the velocity 
offset of the peak \hi\ \21\ optical depth from the strongest metal component in the SDSS spectrum  ($v_{\rm off}$). The \hi\ \21\ 
absorption lines show a wide range of velocity spreads, with $v_{\rm 90}$ $\sim$ 13$-$158\,\kms, and are detected within $\pm$100\,\kms\ 
of the redshift obtained from metal line absorption in the SDSS spectra. 

\begin{figure*}
\subfloat{\includegraphics[height=0.4\textheight, angle=90]{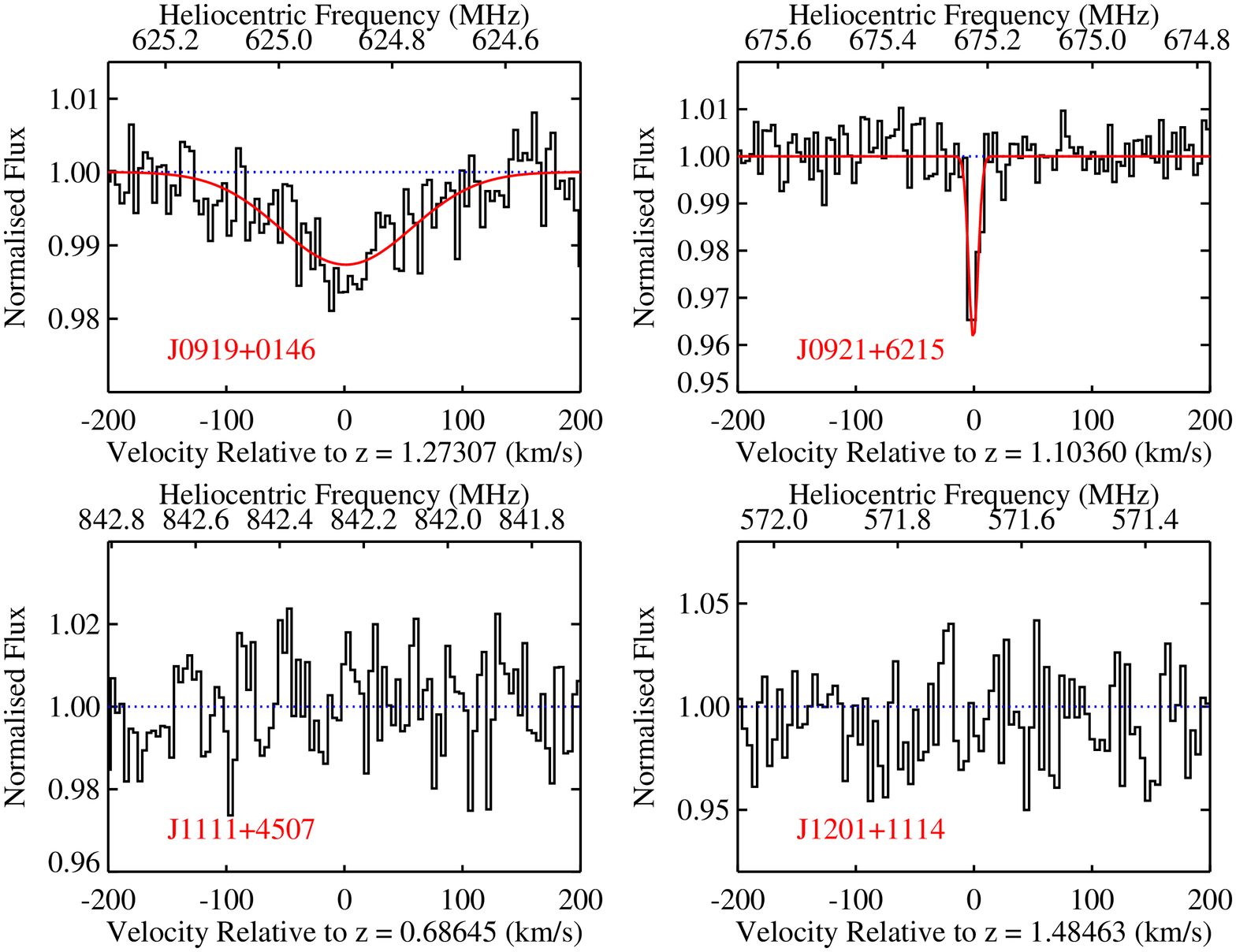} } 
\subfloat{\includegraphics[height=0.4\textheight, angle=90]{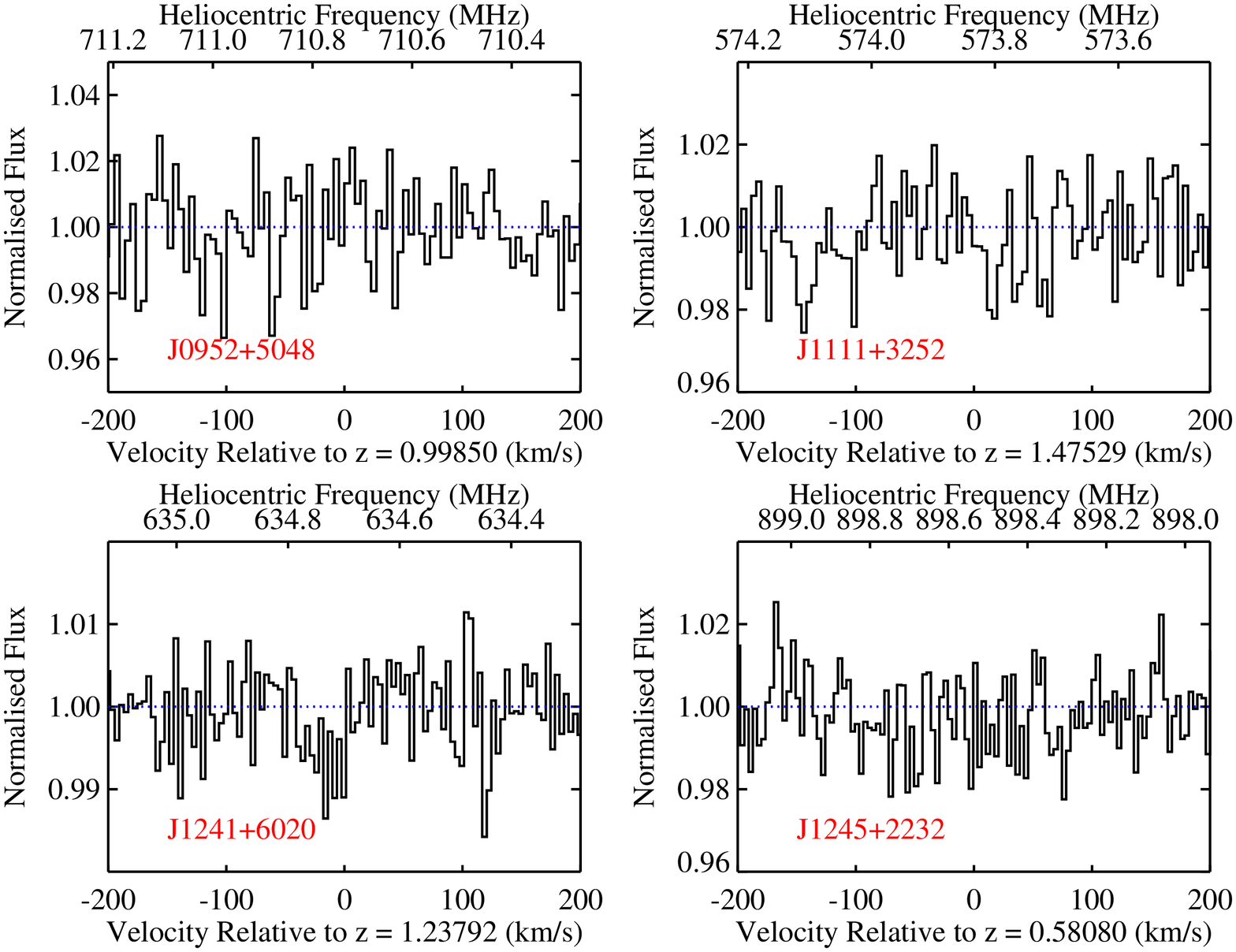} } \hspace{0.01cm}
\subfloat{\includegraphics[height=0.4\textheight, angle=90]{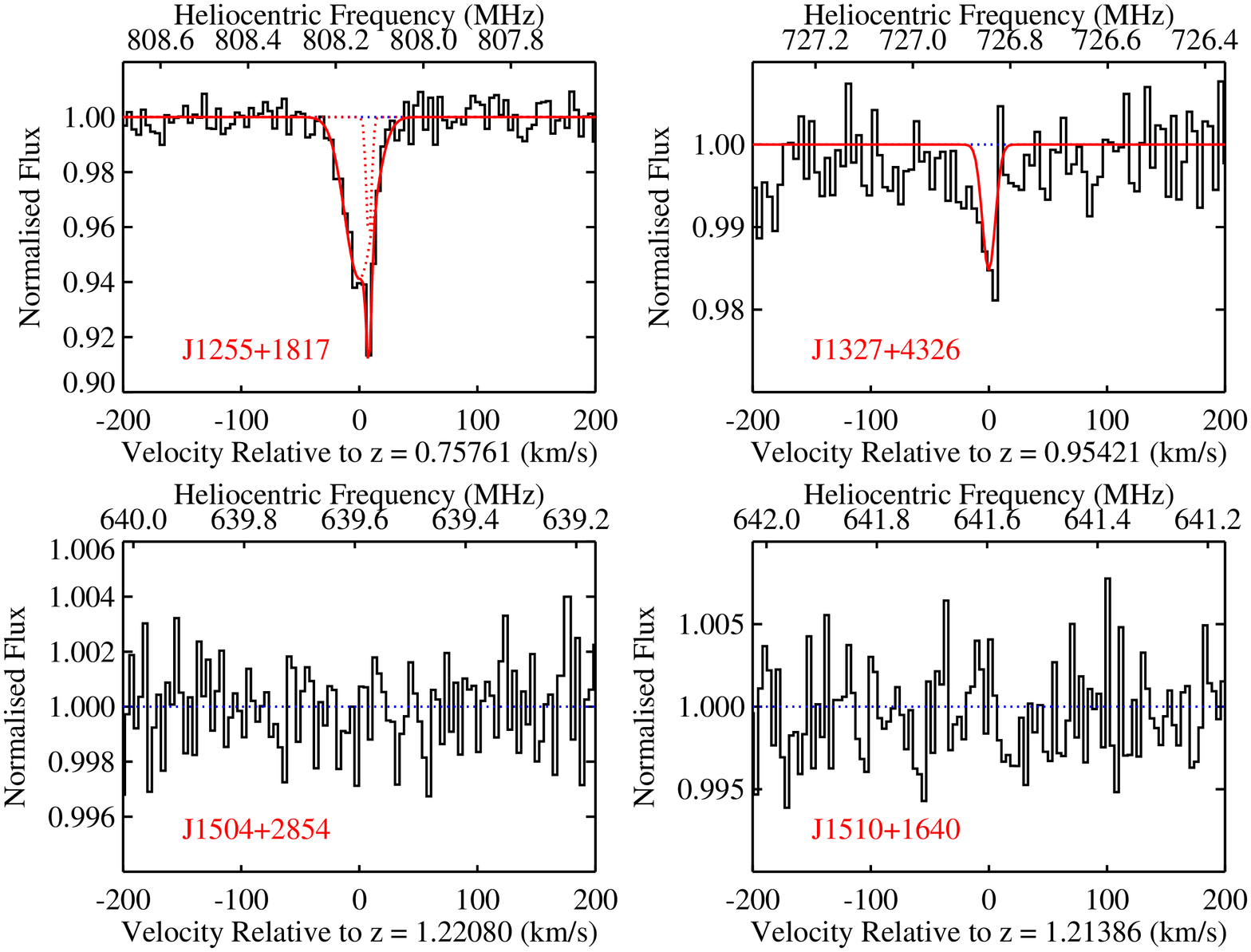} }
\subfloat{\includegraphics[height=0.4\textheight, angle=90]{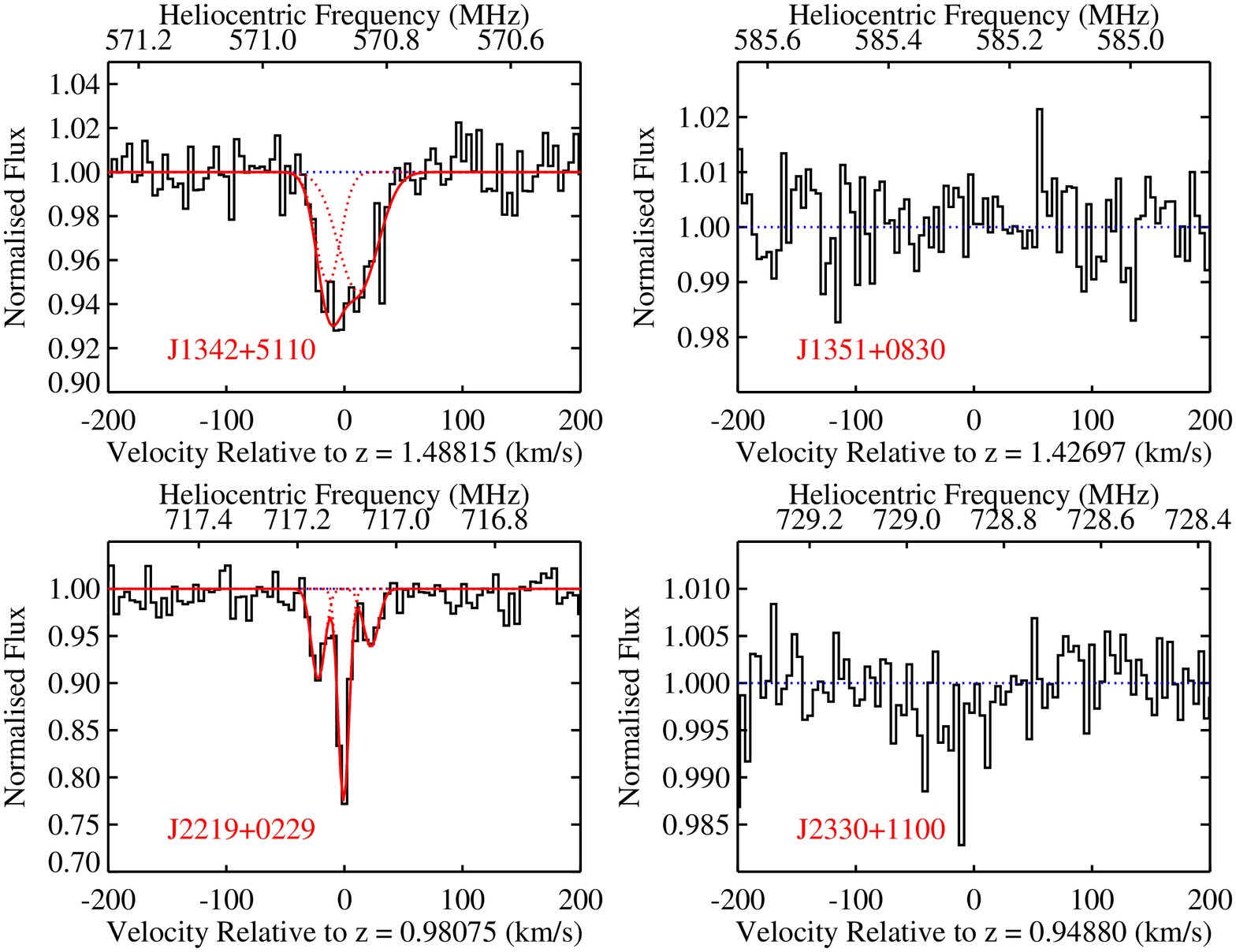} }
\caption{\hi\ \21\ absorption spectra towards the sample of strong \feii\ systems, smoothed to $\sim$4\,\kms. In case of \hi\ \21\ absorption detections, 
individual Gaussian components and the resultant fits to the absorption profiles are overplotted as dotted and continuous lines, respectively. The quasar 
name as given in Table~\ref{tab:sample} is provided for each spectrum.}
\label{fig:21cmspectra1}
\end{figure*}

The \hi\ \21\ absorption spectra are shown in Fig.~\ref{fig:21cmspectra1}. We note that tentative absorption features are present 
towards J1241$+$6020 and J2330$+$1100 at the redshift of the metal absorption. However, these features are not present consistently 
in both the polarizations and are at $\le2\sigma$ significance level. Hence, we consider these as non-detections. The Gaussian fits 
to the \hi\ \21\ absorption profiles are overplotted in Fig.~\ref{fig:21cmspectra1}. The number of Gaussian components is determined 
based on the fit with the minimum \chin. The details of the Gaussian fits, i.e. \zabs, full-width-at-half-maximum (FWHM) and $\tau_{\rm p}$ 
of individual Gaussian components, are provided in Table~\ref{tab:gaussfit}. Constraints on the kinetic temperature, $T_{\rm k}$, using the 
FWHM of the absorption lines, i.e.,
\begin{equation}
 T_{\rm k} < 21.855 \times {\rm FWHM^2~~K,}
\end{equation} 
and on the \nhi, i.e.,
\begin{equation}
 N {\rm (H~\textsc{i})} = 1.93 \times 10^{18}~\tau_{\rm p} \frac{T_{\rm s}}{C_{\rm f}}~{\rm FWHM~~cm^{-2},}
\end{equation}
are provided in the same table.
The FWHMs of the absorption components range from $\sim$5$-$131\,\kms. The large linewidths are most likely dominated by mechanisms 
other than pure thermal broadening, since even the typical temperature ($\sim$10$^4$ K) in WNM corresponds to a FWHM of $\sim$20\,\kms\ 
\citep{wolfire1995,heiles2003}. Assuming a typical \ts\ of 500 K and \fc\ = 1, we find that all the \hi\ \21\ absorption components will 
be arising from DLAs.
\begin{table*} 
\caption{Parameters derived from the redshifted \hi\ \21\ absorption spectra}
\centering
\begin{tabular}{ccccccccccc}
\hline
Quasar & Peak Flux    & $\delta v$ & Spectral     & $\sigma_\tau$ & $\tau_{\rm p}$ & \taudvl\ & \taudv\ & \nhi\             & $v_{\rm 90}$ & $v_{\rm off}$ \\
       & Density      &            & rms          &               &                &          &         & (\ts$/500$ K)     &              &               \\
       & (mJy         & (\kms)     & (mJy         &               &                & (\kms)   & (\kms)  & ($1/$\fc)         & (\kms)       & (\kms)        \\
       & beam$^{-1}$) &            & beam$^{-1}$) &               &                &          &         & ($10^{20}$\,\cms) &              &               \\
(1)    & (2)          & (3)        & (4)          & (5)           & (6)            & (7)      & (8)     & (9)               & (10)         & (11)          \\
\hline
J0919$+$0146 & 183  & 2.0 & 1.0 & 0.005 & 0.02 & $\le$0.068 & 1.61 $\pm$ 0.11 & 15 $\pm$ 1 & 158 & $-$100 \\  
J0921$+$6215 & 1332 & 1.8 & 9.6 & 0.009 & 0.05 & $\le$0.082 & 0.40 $\pm$ 0.05 &  4 $\pm$ 1 & 13  & 13     \\
J0952$+$5048 & 155  & 2.3 & 2.0 & 0.013 & ---  & $\le$0.199 & ---             & $\le$1.8   & --- & ---    \\
J1111$+$3252 & 239  & 2.1 & 3.6 & 0.015 & ---  & $\le$0.188 & ---             & $\le$1.7   & --- & ---    \\
J1111$+$4507 & 184  & 1.9 & 2.7 & 0.015 & ---  & $\le$0.203 & ---             & $\le$1.9   & --- & ---    \\
J1201$+$1114 & 109  & 2.1 & 3.7 & 0.034 & ---  & $\le$0.424 & ---             & $\le$3.9   & --- & ---    \\
J1241$+$6020 & 259  & 1.9 & 2.4 & 0.009 & ---  & $\le$0.150 & ---             & $\le$1.4   & --- & ---    \\
J1245$+$2232 & 180  & 1.8 & 2.1 & 0.012 & ---  & $\le$0.250 & ---             & $\le$2.3   & --- & ---    \\
J1255$+$1817 & 856  & 2.0 & 6.4 & 0.007 & 0.10 & $\le$0.095 & 2.09 $\pm$ 0.08 & 19 $\pm$ 1 & 40  & $-$67  \\
J1327$+$4326 & 647  & 2.3 & 4.4 & 0.007 & 0.02 & $\le$0.080 & 0.36 $\pm$ 0.07 &  3 $\pm$ 1 & 20  & $-$2   \\
J1342$+$5110 & 157  & 2.2 & 2.7 & 0.017 & 0.10 & $\le$0.252 & 3.54 $\pm$ 0.23 & 32 $\pm$ 2 & 56  & 25     \\
J1351$+$0830 & 247  & 2.1 & 2.1 & 0.009 & ---  & $\le$0.129 & ---             & $\le$1.2   & --- & ---    \\
J1504$+$2854 & 1082 & 1.9 & 2.3 & 0.002 & ---  & $\le$0.030 & ---             & $\le$0.3   & --- & ---    \\
J1510$+$1640 & 257  & 1.9 & 0.9 & 0.003 & ---  & $\le$0.041 & ---             & $\le$0.4   & --- & ---    \\ 
J2219$+$0229 & 200  & 2.3 & 3.3 & 0.016 & 0.29 & $\le$0.300 & 5.32 $\pm$ 0.22 & 48 $\pm$ 2 & 58  & 59     \\
J2330$+$1100 & 1745 & 2.2 & 9.1 & 0.005 & ---  & $\le$0.066 & ---             & $\le$0.6   & --- & ---    \\
\hline
\end{tabular}
\label{tab:radiopara}
\begin{flushleft} {\it Notes.}
Column 1: quasar name. Column 2: peak flux density in \mjb\ of the local continuum. 
Column 3: channel width in \kms. Column 4: spectral rms in \mjb. 
Column 5: standard deviation of the \hi\ \21\ optical depth. Column 6: maximum \hi\ \21\ optical depth in case of \hi\ \21\ detections. 
Column 7: 3$\sigma$ upper limit on integrated \hi\ \21\ optical depth from spectra smoothed to 10\,\kms. 
Column 8: integrated \hi\ \21\ optical depth in case of \hi\ \21\ detections.
Column 9: \nhi\ assuming \ts\ = 500 K and \fc\ = 1, in units of $10^{20}$\,\cms\ (3$\sigma$ upper limit in case of non-detections).
Column 10: velocity width which contains 90\% of the total optical depth in case of detections. 
Column 11: velocity offset of the peak \hi\ \21\ optical depth from the strongest metal component in the SDSS spectrum in case of detections. \\
Note that the values in Columns 4 and 5 are at the spectral resolution specified in Column 3.
\end{flushleft}
\end{table*}
\begin{table*} 
\caption{Details of the Gaussian fits to the \hi\ \21\ absorption profiles}
\centering
\begin{tabular}{cccccccc}
\hline
Quasar & ID  & \zabs\ & FWHM   & $\tau_{\rm p}$ & $T_{\rm k}$ & \nhi\                   \\
       &     &        &        &                &             & (\ts$/500$ K) ($1/$\fc) \\
       &     &        & (\kms) &                & (K)         & ($10^{20}$\,\cms)       \\
(1)    & (2) & (3)    & (4)    & (5)            & (6)         & (7)                     \\
\hline
J0919$+$0146 & 1 & 1.27308 & 131 $\pm$ 21 & 0.013 $\pm$ 0.004 & $<$373774 & 16 $\pm$ 7  \\
J0921$+$6215 & 1 & 1.10360 & 9   $\pm$ 2  & 0.04 $\pm$ 0.01  & $<$1908   &  4 $\pm$ 2  \\
J1255$+$1817 & 1 & 0.75761 & 29  $\pm$ 2  & 0.06 $\pm$ 0.01  & $<$18060  & 17 $\pm$ 4  \\ 
             & 2 & 0.75766 & 5   $\pm$ 1  & 0.05 $\pm$ 0.03  & $<$591    &  3 $\pm$ 2  \\ 
J1327$+$4326 & 1 & 0.95421 & 13  $\pm$ 7  & 0.02 $\pm$ 0.01  & $<$3600   &  3 $\pm$ 2  \\
J1342$+$5110 & 1 & 1.48804 & 24  $\pm$ 5  & 0.05 $\pm$ 0.04  & $<$12777  & 12 $\pm$ 10 \\
             & 2 & 1.48825 & 39  $\pm$ 10 & 0.06 $\pm$ 0.04  & $<$33987  & 22 $\pm$ 20 \\         
J2219$+$0229 & 1 & 0.98060 & 13  $\pm$ 1  & 0.10 $\pm$ 0.01  & $<$3923   & 13 $\pm$ 3  \\
             & 2 & 0.98075 & 10  $\pm$ 1  & 0.26 $\pm$ 0.02  & $<$2394   & 26 $\pm$ 3  \\
             & 3 & 0.98090 & 15  $\pm$ 2  & 0.06 $\pm$ 0.01  & $<$5023   & 10 $\pm$ 3  \\      
\hline
\end{tabular}
\label{tab:gaussfit}
\begin{flushleft} {\it Notes.}
Column 1: quasar name. Column 2: absorption component identification. Column 3: redshift of absorption component. Column 4: FWHM (\kms) of the Gaussian component. 
Column 5: peak optical depth of the Gaussian component. Column 6: upper limit on $T_{\rm k}$ (K), obtained assuming the line width is purely due to thermal motions. 
Column 7: \nhi\ assuming \ts\ = 500 K and \fc\ = 1, in units of $10^{20}$\,\cms.
\end{flushleft}
\end{table*}
\section{Incidence of \hi\ \21\ absorption in strong \feii\ systems}
\label{sec_coveringfactor}
We define the detection rate of \hi\ \21\ absorption ($C_{21}$) as the fraction of systems showing \hi\ \21\ detections with \taudvl\ $\le$ \t0\ and \taudv\ $\ge$ \t0, where \t0\ is a
3$\sigma$ optical depth sensitivity. We use \t0\ = 0.3 \kms\ (corresponding to a sensitivity of \nhi\ $\le$ 5 $\times$ 10$^{19}$ \cms\ for \ts\ = 100 K and \fc\ = 1) throughout this 
work to estimate $C_{21}$, unless otherwise mentioned \footnote{\t0\ = 0.3 \kms\ is chosen also to enable comparisons with existing \hi\ \21\ absorption measurements in the literature.}. 
The quoted errors represent Gaussian 1$\sigma$ confidence intervals computed using tables of \citet{gehrels1986} assuming a Poisson distribution. Note that due to lack of uniform 
measurement of \fc\ for all the systems, we do not correct the optical depths for \fc\ in subsequent discussions. 

\begin{figure}
\includegraphics[height=0.35\textheight, angle=90]{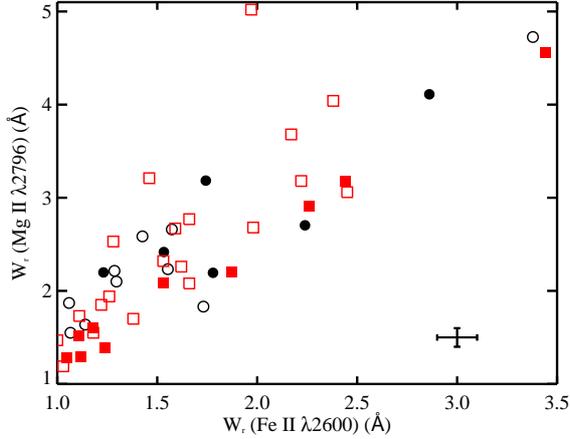} 
\caption{\wmg\ versus \wfe\ for the strong \feii\ absorbers in our sample are shown as black circles and in the systems from G09 and G12 are shown as red squares.
The filled symbols correspond to \hi\ \21\ detections and the open symbols correspond to non-detections. The typical error bars are plotted in the bottom right corner.}
\label{fig:wmgwfe}
\end{figure}

We combine our \hi\ \21\ measurements with those of strong \mgii\ systems (i.e., \wmg\ $\ge$ 1 \AA) at 0.5$<z<$1.5 from G09 and G12, that satisfy our criterion of being strong \feii\ 
absorbers with \wfe\ $\ge$ 1.0\,\AA. Note that for the absorber at \zabs\ = 0.6216 towards J093035.08$+$464408.7 from G12, we consider the recent measurement of \hi\ \21\ absorption 
(which appeared as a tentative feature in G12) using deeper observations by \citet{zwaan2015}. The summary of all the strong \feii\ absorbers considered here for statistical analyses 
is given in Table~\ref{tab:samples}. In total we consider 46 strong \feii\ systems out of which 16 show \hi\ \21\ absorption. We refer to the sample presented in this work (see 
Table~\ref{tab:sample}) as S1, and the combined sample as S2. Note that sample S2 is homogeneous with regards to analysis of the radio data. Fig.~\ref{fig:wmgwfe} shows the distribution 
of \wmg\ and \wfe\ in sample S2.

\begin{figure}
\includegraphics[height=0.35\textheight, angle=90]{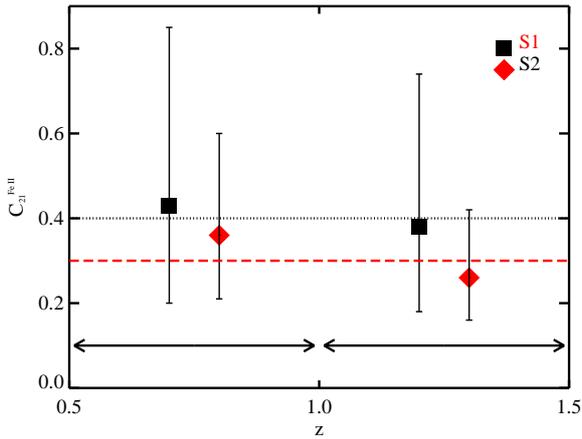}
\caption{The detection rate of \hi\ \21\ absorption (for \t0\ = 0.3\,\kms) in samples S1 (square) and S2 (diamond) over the two redshift ranges, 0.5$<z<$1.0 and 1.1$<z<$1.5 (demarcated by arrows). 
The dotted and dashed horizontal lines represent the mean \c21\ in samples S1 and S2, respectively, over 0.5$<z<$1.5 (see Section~\ref{sec_coveringfactor}).}
\label{fig:coveringfactor1}
\end{figure}

\begin{table}  
\centering
\caption{Summary of strong \feii\ absorbers at 0.5$<z<$1.5 with \hi\ \21\ absorption measurements considered here}
\begin{tabular}{lccc}
\hline
Sample & No. of  & \hi\ \21\  & Reference  \\
       & systems & detections &            \\
\hline
S1  & 16 & 6     & This work          \\
G09 & 18 & 5     & \citet{gupta2009}  \\
G12 & 12 & 5$^*$ & \citet{gupta2012}  \\
S2  & 46 & 16    & S1 $+$ G09 $+$ G12 \\
\hline
\end{tabular}
\label{tab:samples}
\begin{flushleft}
$^*$ With the improved parameters from the \hi\ \21\ absorption spectrum towards J093035.08$+$464408.7 reported in \citet{zwaan2015}
\end{flushleft}
\end{table}
We estimate the detection rate of \hi\ \21\ absorption in the \feii\ samples over 0.5$<z<$1.5 as, \c21\ = 0.40$^{+0.24}_{-0.16}$ (0.36$^{+0.29}_{-0.17}$ for \t0\ = 0.2\footnote{Since 
lower values of \t0\ will select the weaker \hi\ 21-cm absorbers and higher values will select the stronger absorbers, we provide the detection rates for \t0\ = 0.2 for comparison.}) 
for S1, and \c21\ = 0.30$^{+0.12}_{-0.09}$ (0.39$^{+0.16}_{-0.12}$ for \t0\ = 0.2) for S2. We observe no evolution of \c21\ within the uncertainties over the redshift range studied, 
for both S1 and S2 (see Fig.~\ref{fig:coveringfactor1}). This is consistent with the constant detection rate of \hi\ \21\ absorption in strong \mgii\ systems found by G12. 
Recall that our sample selection procedure picked three strong \feii\ systems that are part of the \mgii\ sample of K09 and one system that is part of the \mgii\ sample of 
L00 (see Section~\ref{sec_sample}). Including these measurements from K09 and L00 in the sample S2 does not change our statistical results within the uncertainties, 
i.e. \c21\ = 0.27$^{+0.11}_{-0.08}$ (0.39$^{+0.15}_{-0.11}$ for \t0\ = 0.2). 

\begin{figure}
\subfloat[]{\includegraphics[height=0.35\textheight, angle=90]{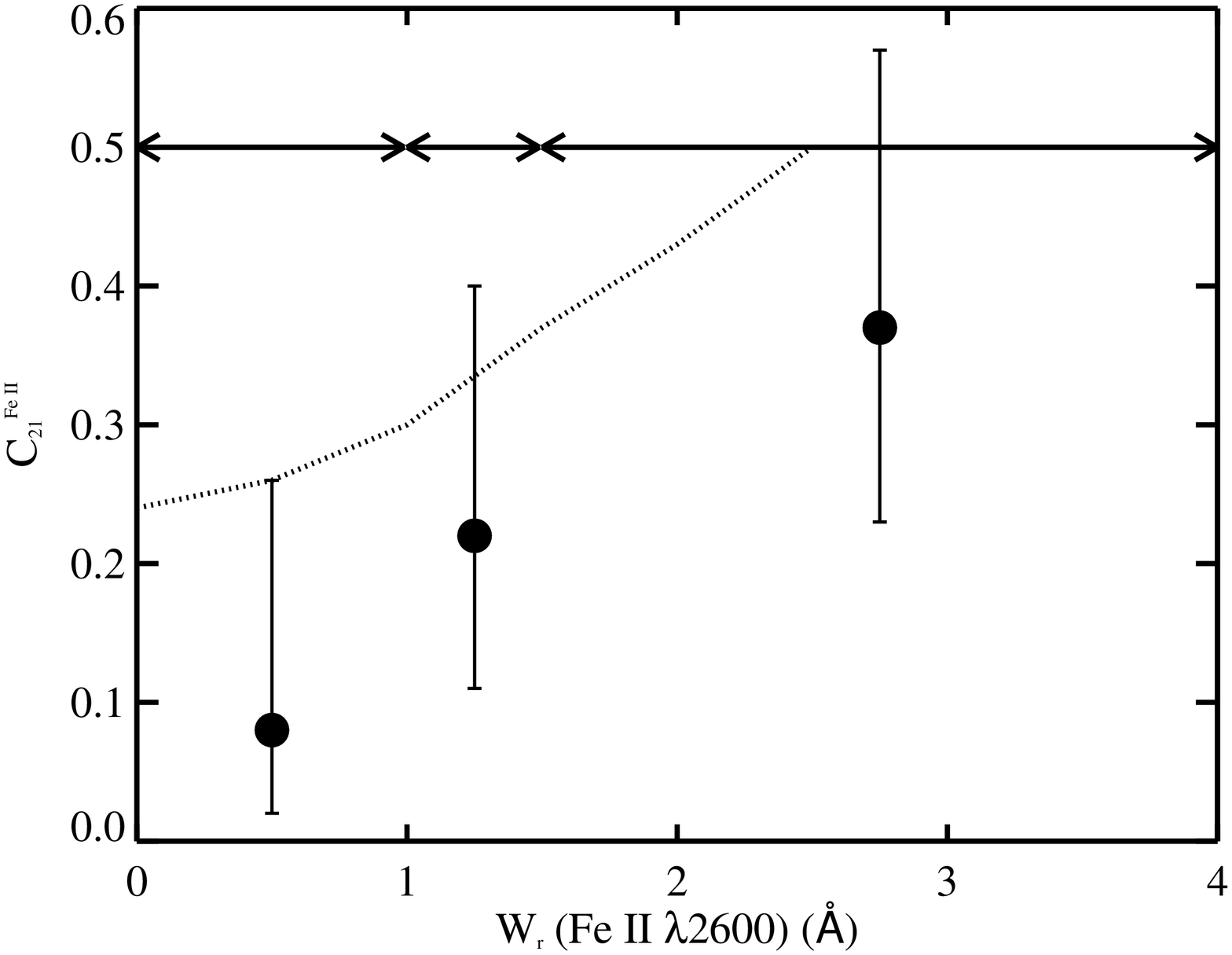} } 
\caption{The detection rate of \hi\ \21\ absorption (for \t0\ = 0.3\,\kms) in three different bins (demarcated by arrows) of \wfe\ for sample S2. The dotted line is the cumulative distribution of \c21, 
i.e. for \wfe\ = $x$ \AA, it gives \c21\ in systems with \wfe\ $\ge$ $x$ \AA. The \c21\ estimate at \wfe\ $<$ 1 \AA\ are using measurements from G09 and G12.}
\label{fig:coveringfactor2}
\end{figure}

To determine the detection efficiency of \hi\ \21\ absorption of a \feii\ equivalent width based selection technique, we plot \c21\ as a function of \wfe\ in Fig.~\ref{fig:coveringfactor2}.
For the strong \feii\ systems in S2, we plot \c21\ in two bins of \wfe\ demarcated at the median value, \wfe\ = 1.5 \AA. For comparison, we also plot \c21\ (0.08$^{+0.18}_{-0.06}$) in strong 
\mgii\ systems with \wfe\ $<$ 1 \AA\ from G09 and G12. We caution that the strong \mgii\ systems, that have been searched for \hi\ \21\ absorption in the literature, are biased towards 
high \wfe, with $\sim$65\% of them having \wfe\ $\ge$ 1.0, compared to 37\% of the strong \mgii\ systems in SDSS at 0.5$<z<$1.5 from \citet{zhu2013} having \wfe\ $\ge$ 1.0. Hence, the number 
of systems with \wfe\ $<$ 1 \AA\ is only one third of sample S2. We find that the detection efficiency of \hi\ \21\ absorption in systems with \wfe\ $\ge$ 1 \AA\ (sample S2) is a factor of 
$\sim$3.8 times higher than that in systems with \wfe\ $<$ 1 \AA, albeit not statistically significant due to the large uncertainties. However, the increasing trend of \c21\ with \wfe\ is 
valid over both the redshift ranges, 0.5$<z<$1.0 and 1.1$<z<$1.5. 

Additionally, we plot in Fig.~\ref{fig:coveringfactor2} the cumulative distribution of \c21\ as a function of \wfe, i.e., for \wfe\ = $x$ \AA, it gives \c21\ in systems with \wfe\ $\ge$ $x$ \AA. 
It can be seen that \c21\ increases as we go for samples with higher \wfe. This increasing trend also holds for different optical depth sensitivities, on correcting the optical depths for \fc, 
and on considering the additional measurements from K09 and L00. G09 and G12 have found that the detection rate of \hi\ \21\ absorption can be increased by up to a factor of 2 by imposing constraints 
on DR, R1 and R2. Here we have demonstrated that a simple \wfe\ $\ge$ 1 \AA\ based selection would work just as well for increasing the probability of detecting \hi\ \21\ absorption. Additionally, 
we note that assuming a typical \ts\ = 500 K (see Section \ref{sec_parameters}) and \fc\ = 1, all the 16 \hi\ \21\ absorption detected in strong \feii\ systems would be arising from DLAs. Hence, 
strong \feii\ systems seem to have a high probability of harbouring high \nhi\ cold gas.

From the detection rate of \hi\ \21\ absorption in strong \feii\ systems and that of DLAs in strong \feii\ systems, we can estimate the detection rate of \hi\ \21\ absorption in DLAs,
i.e., $C_{21}^{\rm DLA} = C_{21}^{\rm Fe~\textsc{ii}} / C_{\rm DLA}^{\rm Fe~\textsc{ii}}$. Using $C_{21}^{\rm Fe~\textsc{ii}}$ obtained for S2 (see Fig.~\ref{fig:coveringfactor1}) and 
$C_{\rm DLA}^{\rm Fe~\textsc{ii}}$ = 0.45$^{+0.10}_{-0.08}$ obtained for the $z<$1.65 \mgii\ sample of R06 (see Section~\ref{sec_sample}), we estimate $C_{21}^{\rm DLA}$ = 0.80$^{+0.20}_{-0.36}$ 
at 0.5$<z<$1.0 and $C_{21}^{\rm DLA}$ = 0.58$^{+0.38}_{-0.25}$ at 1.1$<z<$1.5. In comparison, the detection rate of \hi\ \21\ absorption in $z<$1 DLAs is 0.62$^{+0.30}_{-0.21}$ \citep[see][]{dutta2016b}, 
and that in 2.0$\le z\le$3.5 DLAs is 0.22$^{+0.29}_{-0.14}$ \citep[see][]{srianand2012}. Note that all the detection rates estimates are for \t0\ = 0.3\,\kms. Hence, $C_{21}^{\rm DLA}$ 
appears to be increasing with decreasing redshift, even though the errors are large. This increasing trend of the \hi\ \21\ absorption detection rate in DLAs with decreasing redshift has 
also been noted by K09 and G12 \citep[however see also,][]{curran2005,curran2006,curran2010b,curran2012}. This is similar to the increase in detection rate of \h2\ absorption in DLAs/sub-DLAs 
from $\sim$10\% at $z\ge1.8$ \citep{noterdaeme2008} to $\sim$50\% at $z\le0.7$ \citep{muzahid2015}. In addition, as noted by \citet{dutta2016b}, the incidence of \hi\ \21\ absorption in 
absorption-selected samples is much higher than that in $z<0.4$ quasar-galaxy pairs that are selected on the basis of galaxies, implying patchy distribution and small sizes (parsec to 
sub-parsec scale) of cold gas clouds in the extended discs/haloes of galaxies. The increase in incidence of \hi\ \21\ absorption with decreasing redshift can be explained as a result of 
the increase in the CNM filling factor in galaxies with time. 

\section{\hi\ \21\ absorption and metals}
\label{sec_metals}
\subsection{Metal line properties}
\label{sec_metals1}
\begin{figure*}
\subfloat[]{\includegraphics[height=0.25\textheight, angle=90]{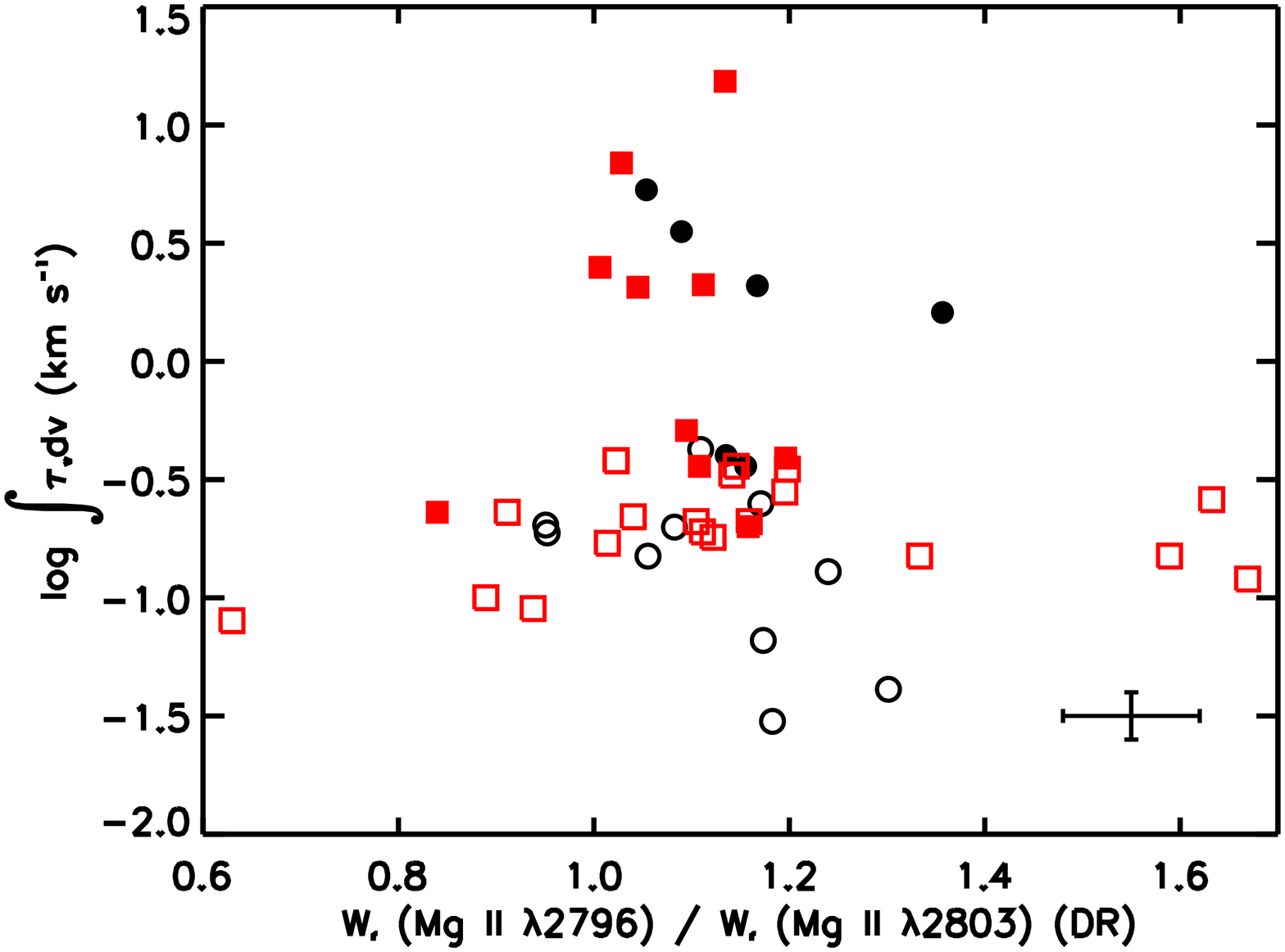} } 
\subfloat[]{\includegraphics[height=0.25\textheight, angle=90]{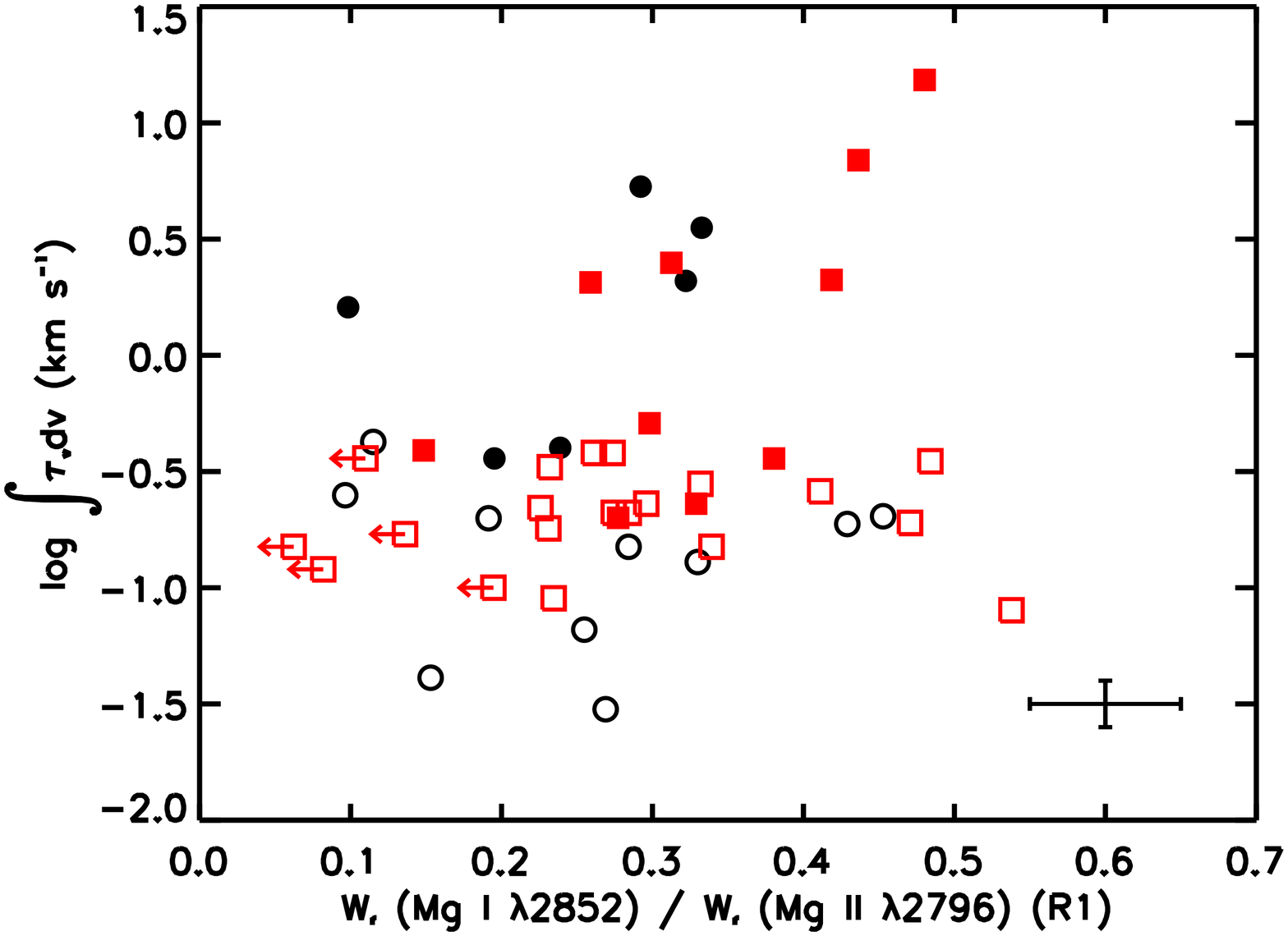} }
\subfloat[]{\includegraphics[height=0.25\textheight, angle=90]{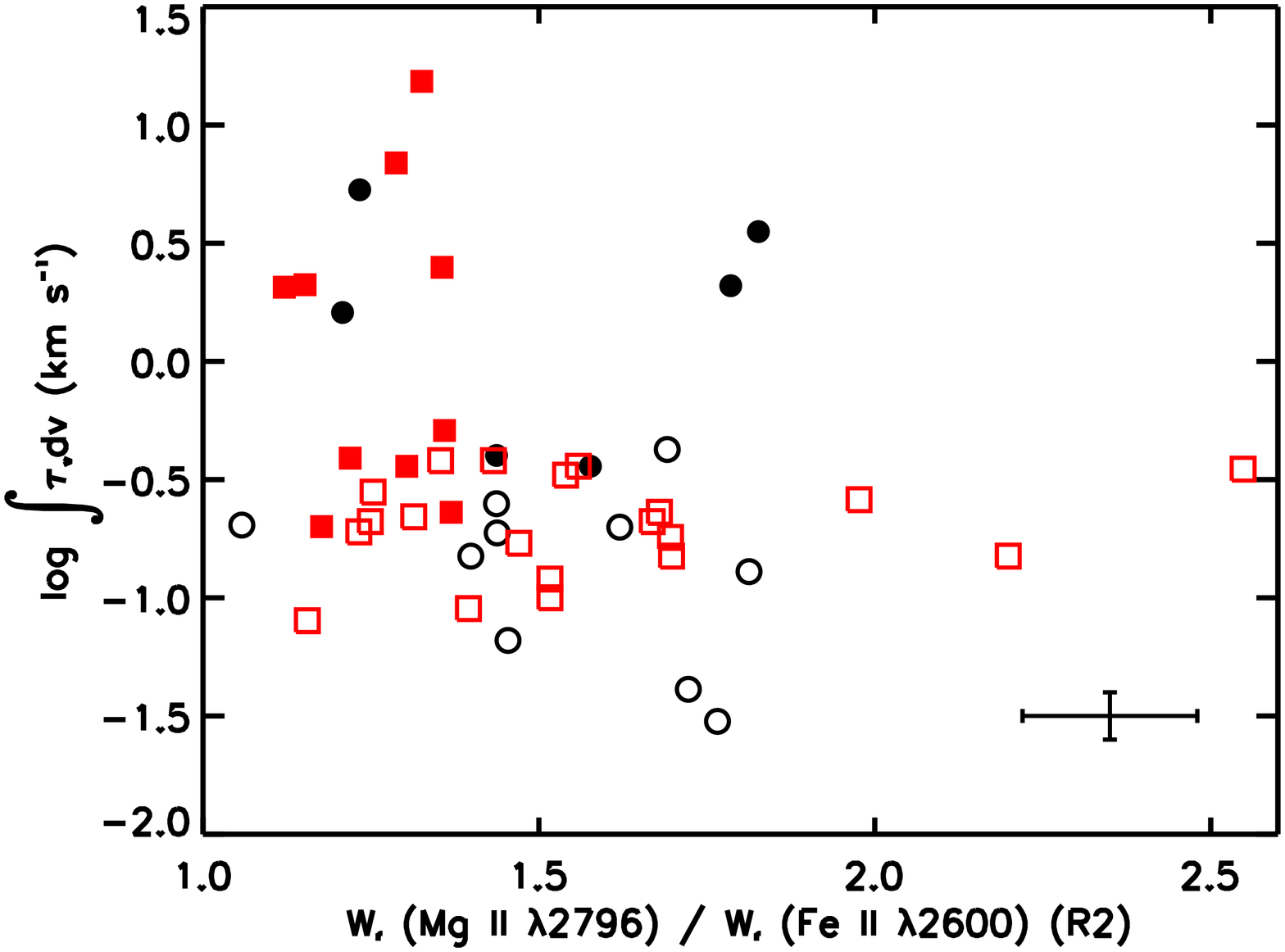} }
\caption{\hi\ \21\ absorption strength of the strong \feii\ systems as a function of different metal line equivalent width ratios: (a) DR, (b) R1, and (c) R2. 
Black circles are for the systems in our sample and red squares are for the systems from G09 and G12. 
The filled symbols correspond to \taudv\ of the \hi\ \21\ detections and the open symbols correspond to \taudvl\ of the non-detections.
The typical error bars are plotted in the bottom right corner of each plot.}
\label{fig:metalratios}
\end{figure*}

The metal line ratios like DR, R1 and R2 are found to be more robust indicators of high \nhi\ systems than \wmg\ alone (R06). Further, G09 and G12 have found that a higher
detection rate of \hi\ \21\ absorption can be obtained in systems limited to restricted ranges of these parameters. Here, we investigate the dependence of \hi\ \21\ absorption 
in strong \feii\ systems on their metal absorption. For the literature systems, we consider the rest equivalent widths of metal lines as given in G09 and G12. We plot the 
\hi\ \21\ absorption strength of the strong \feii\ systems in S2 as a function of the metal line equivalent width ratios, DR, R1 and R2 in Fig.~\ref{fig:metalratios}. We find a 
positive correlation between \taudv\ of the \hi\ \21\ absorption detections and R1: Kendall's rank correlation coefficient, $r_{\rm k}$ = 0.38, with the probability of the correlation 
arising by chance, $P(r_{\rm k})$ = 0.04, which is significant at $S(r_{\rm k})$ = 2.0$\sigma$ assuming Gaussian statistics; and Spearman rank correlation coefficient, $r_{\rm s}$ = 0.49, 
$P(r_{\rm s})$ = 0.05, $S(r_{\rm s})$ = 1.9$\sigma$. Since higher R1 values are likely to be arising in systems with high \nhi, this may imply a weak correlation of \taudv\ with \nhi. 
However, the correlation becomes less significant ($1.4\sigma$) when we include the upper limits (\taudvl) as censored data points and perform survival analysis using the 
{\sc `cenken'} function under the {\sc `nada'} package in {\sc r}. We also do not find any correlation of the \hi\ \21\ absorption strength with \wmg, \wfe, DR and R2 
(i.e. they are at $\lesssim1\sigma$ significance). Note that most of the \hi\ \21\ detections arise in systems with DR $\sim$1, i.e. in the saturated regime.

It is important to keep in mind that the metal line equivalent widths and ratios measured from low resolution SDSS spectra are averaged over several absorption components. 
High resolution optical spectra of \mgii\ systems show that the metal line absorption usually consists of multiple components, with the number of components increasing 
with \wmg\ \citep[e.g.][]{churchill2003}. Hence, the lack of correlation between the \hi\ \21\ absorption strength and the metal line parameters estimated from SDSS could 
be because the \hi\ \21\ absorption arises from certain specific metal line components \citep[e.g. G09;][]{srianand2012,rahmani2012,dutta2015}. This emphasizes the need for 
high resolution optical spectra of the metal absorption lines. 

Finally, there are eleven ultrastrong \mgii\ systems, i.e. those with \wmg\ $\ge$ 3.0\,\AA, in S2. It has been suggested that ultrastrong \mgii\ absorbers may arise in galactic superwinds 
\citep{bond2001,nestor2011}. However, their large equivalent widths could also be driven by gas dynamics of the intragroup medium \citep[e.g.][]{gauthier2013}. Typically $\sim$7\% of \wmg\ 
$\ge$ 1 \AA\ absorbers in SDSS satisfy the definition of ultrastrong \mgii\ absorbers. This fraction becomes $\sim$12\% when the absorbers also have \wfe\ $\ge$ 1 \AA. However, ultrastrong \mgii\ 
systems constitute $\sim$24\% of sample S2. \hi\ \21\ absorption has been detected in four of these. We do not find these absorbers to be different from the rest of the strong \feii\ absorbers 
in terms of their \hi\ \21\ absorption detection rate, optical depths and velocity widths. While the metal equivalent widths are large in these systems, the velocity widths of the \hi\ \21\ 
absorption are in the range of $\sim$12$-$60\,\kms.
\subsection{Average metal content}
\label{sec_metals2}
\begin{figure*}
\subfloat[]{\includegraphics[width=0.45\textwidth, angle=90, clip]{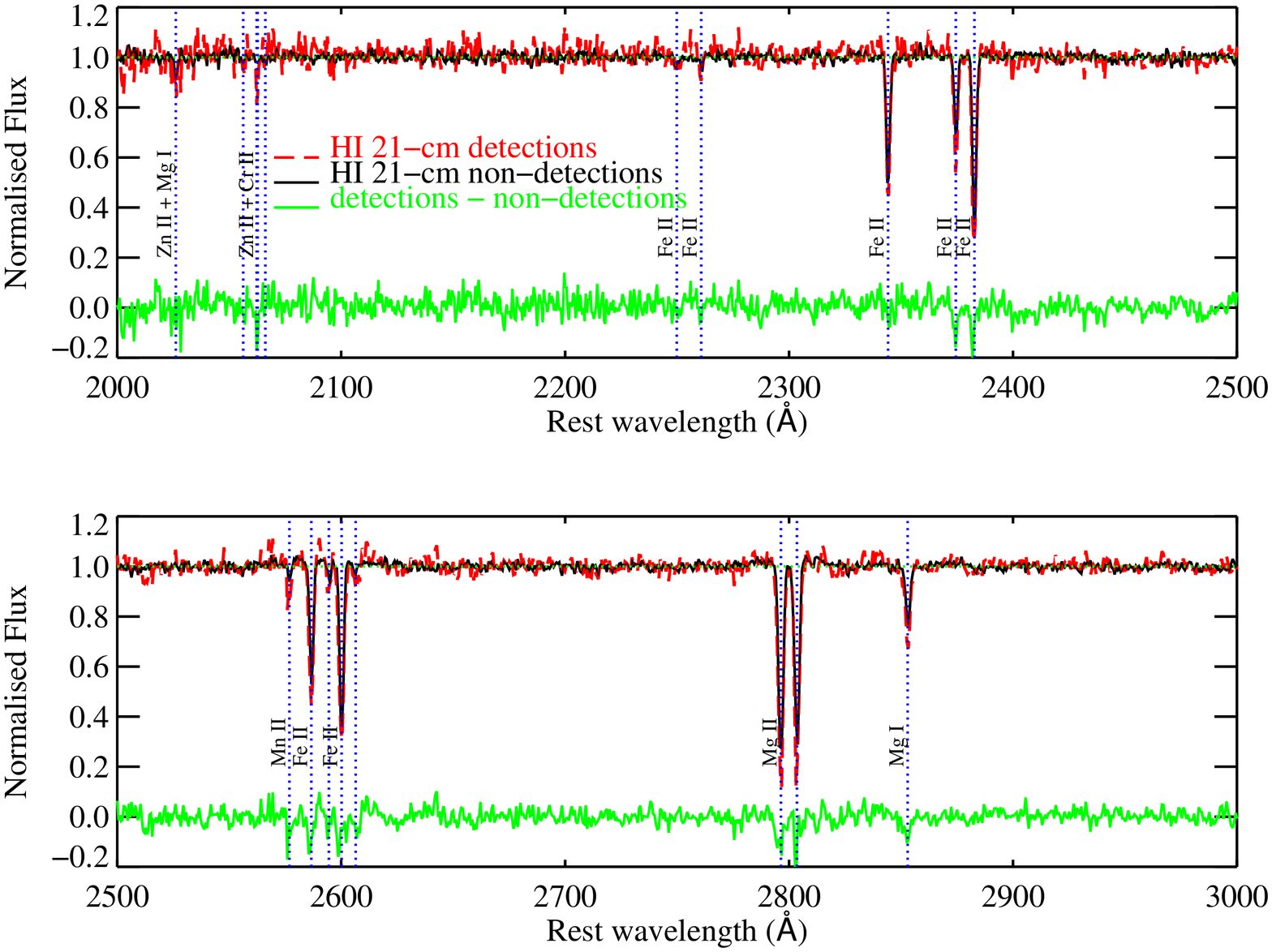}}
\subfloat[]{\includegraphics[height=0.3\textheight, angle=90]{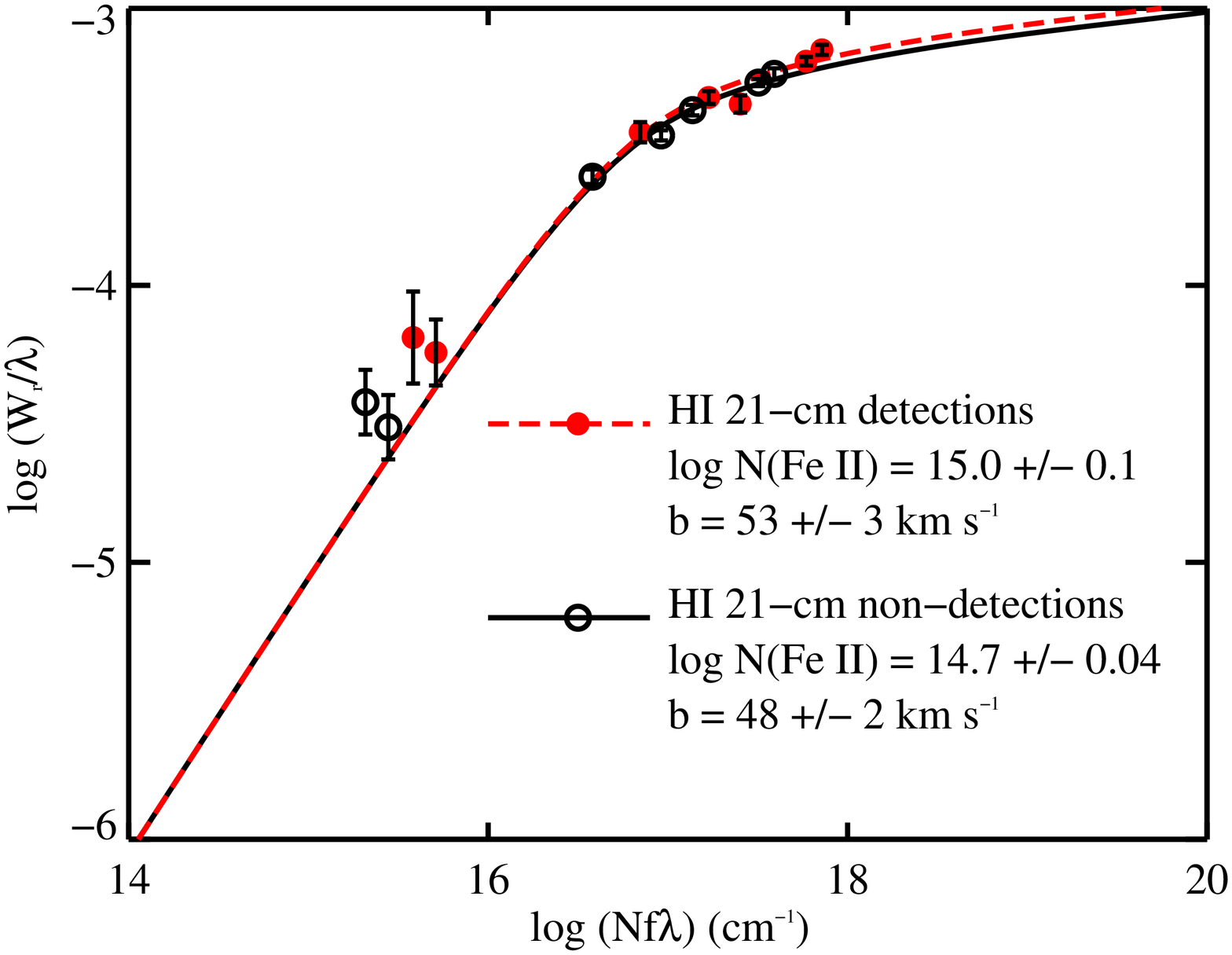}}
\caption{(a) The median stacked spectrum of SDSS quasars with strong \feii\ systems at 0.5$<z<$1.5 (in rest frame) which are detected in \hi\ \21\ absorption is shown as the red dashed line, 
and the same for systems which are not detected in \hi\ \21\ absorption is shown as the black solid line. The difference of the stacked spectrum of \hi\ \21\ non-detections from that of the 
detections is shown as the green line at an arbitrary offset in the y-axis for clarity. The various rest wavelengths of transitions of \mgii, \mgi, Cr\,{\sc ii}, Mn\,{\sc ii}, \feii\ and 
Zn\,{\sc ii} are marked by vertical dotted lines. (b) The curve-of-growth for \feii\ lines detected in the stacked spectrum of \hi\ \21\ detections is shown as the red dashed line and that 
for \feii\ lines detected in the stacked spectrum of \hi\ \21\ non-detections is shown as the solid black line. The filled and open circles mark the observed rest-frame equivalent widths 
and the column densities calculated for \feii\ lines in the stacked spectrum of \hi\ \21\ detections and non-detections, respectively.}
\label{fig:stack}
\end{figure*}

To check if there is any difference in the average metal line properties of the systems which show \hi\ \21\ absorption from those which do not, we stacked the SDSS spectra of the background quasars
in sample S2, with and without \hi\ \21\ absorption separately. The resultant median stacked spectra are shown in panel (a) of Fig.~\ref{fig:stack}, and the rest equivalent widths of different 
metal transitions in the stacked spectra are listed in Table~\ref{tab:metalstack}. We consider the rest wavelength range of 2000$-$3000\,\AA, which is uniformly covered for all the systems. It can 
be seen that the systems which show \hi\ \21\ absorption, also show systematically stronger absorption (i.e. larger equivalent widths by $\sim3-4\sigma$) from \mgii, \mgi\ and \feii, than those which 
do not. The absorption lines of Cr\,{\sc ii}, Mn\,{\sc ii} and Zn\,{\sc ii} are also stronger for systems with \hi\ \21\ detections than for those without, though the differences in equivalent widths 
are not more that $\sim2\sigma$ significant. We expect strong \feii\ and \hi\ \21\ absorption to probe sightlines close to galaxies. Nebular emission lines are usually detected in the stacked spectra of 
strong \mgii\ absorbers \citep{noterdaeme2010b,menard2011}. However, the number of systems in our stacked spectrum of \hi\ \21\ absorbers is small, and $3\sigma$ upper limits on average luminosity 
of the nebular emission lines of H$\beta$, [O\,{\sc ii}] and [O\,{\sc iii}] are 5, 4 and 5, in units of $10^{40}$ erg~s$^{-1}$, respectively.

Using different absorption lines of \feii\ detected in the stacked spectra, we compute the column density ($N$) and effective Doppler parameter ($b$) by constructing a single cloud curve-of-growth (COG)
(see panel (b) of Fig.~\ref{fig:stack}). Note that the spectral resolution of the SDSS spectra does not allow us to measure $N$ and $b$ accurately. However, a single cloud COG can still give an indication 
of the column density and velocity field. The effective $b$ parameter obtained for \feii\ lines in the stacked spectrum of \hi\ \21\ detections is slightly higher than that for \feii\ lines in the stacked 
spectrum of non-detections. The larger effective $b$ parameter could reflect larger number of absorption components, and hence higher probability of detecting \hi\ \21\ absorption from a cold gas component. 
The stacked spectrum of \hi\ \21\ detections gives two times higher $N$(\feii) than that of the non-detections. Even if we consider only the two weak transitions of \feii\ $\lambda$2249 and $\lambda$2260, 
which are unlikely to be affected by saturation, we get two times higher $N$(\feii) for the stacked spectrum of \hi\ \21\ detections compared to that of the non-detections. Therefore, it seems that systems 
which give rise to \hi\ \21\ absorption are likely to have stronger metal absorption (i.e. higher column density) on average than those which do not. This is also consistent with the increase in detection 
rate of \hi\ \21\ absorption with increasing \wfe\ (see Fig.~\ref{fig:coveringfactor2}).
\begin{table} 
\caption{List of rest equivalent widths of different metal transitions in the stacked spectra discussed in Section~\ref{sec_metals2}}
\centering
\begin{tabular}{cccc}
\hline
Transition & REW(detections) & REW(non-detections) \\
           & (\AA)           & (\AA)               \\        
(1)        & (2)             & (3)                 \\
\hline
\mgii\ $\lambda$2796 & 2.22 $\pm$ 0.08 & 1.88 $\pm$ 0.05 \\
\mgii\ $\lambda$2803 & 2.08 $\pm$ 0.07 & 1.75 $\pm$ 0.05 \\
\mgi\  $\lambda$2852 & 0.77 $\pm$ 0.07 & 0.54 $\pm$ 0.03 \\
\feii\ $\lambda$2600 & 1.68 $\pm$ 0.06 & 1.40 $\pm$ 0.04 \\
\feii\ $\lambda$2586 & 1.23 $\pm$ 0.06 & 0.90 $\pm$ 0.04 \\
\feii\ $\lambda$2382 & 1.69 $\pm$ 0.07 & 1.38 $\pm$ 0.06 \\
\feii\ $\lambda$2374 & 0.85 $\pm$ 0.07 & 0.59 $\pm$ 0.03 \\
\feii\ $\lambda$2344 & 1.06 $\pm$ 0.04 & 1.01 $\pm$ 0.02 \\
\feii\ $\lambda$2260 & 0.13 $\pm$ 0.04 & 0.07 $\pm$ 0.02 \\
\feii\ $\lambda$2249 & 0.15 $\pm$ 0.04 & 0.09 $\pm$ 0.02 \\
\mnii\ $\lambda$2606 & 0.14 $\pm$ 0.06 & 0.05 $\pm$ 0.04 \\
\mnii\ $\lambda$2594 & 0.20 $\pm$ 0.06 & 0.06 $\pm$ 0.04 \\
\mnii\ $\lambda$2576 & 0.28 $\pm$ 0.06 & 0.07 $\pm$ 0.04 \\
\znii\ $+$ \crii\ $\lambda$2062 & 0.22 $\pm$ 0.06 & 0.03 $\pm$ 0.03 \\
\znii\ $+$ \mgi\ $\lambda$2026  & 0.25 $\pm$ 0.06 & 0.13 $\pm$ 0.02 \\
\hline
\end{tabular}
\label{tab:metalstack}
\begin{flushleft} {\it Notes.}
Column 1: metal line transition. Columns 2 and 3: rest equivalent widths of the transition in the stacked spectra of quasars with \hi\ \21\ absorption detections and non-detections, respectively. \\
\end{flushleft}
\end{table}
\section{Dust content of \hi\ \21\ absorbers}
\label{sec_dust}
Dust grains play an important role in the heating and cooling of the neutral gas phases of the ISM \citep{wolfire1995}, and also in the formation of \h2\ molecules in the cold
and dense phases \citep{gould1963,hollenbach1971}. The large homogeneous database of SDSS quasar spectra has made it possible to statistically quantify the dust content of absorber 
populations like \mgii\ systems, Ca\,{\sc ii} systems and DLAs through differential reddening measurements \citep{york2006,wild2006,menard2008,vladilo2008,budzynski2011,khare2012,menard2012,sardane2015,murphy2016}. 
Such studies have found that \mgii\ systems rarely show strong reddening due to dust in the spectra of the background quasars (mean \ebv\ $\sim$ 0.02). \citet{jiang2011} have 
found only 39 cases that are 2175 \AA\ dust extinction bump candidates among 2951 strong \mgii\ systems from SDSS with \wmg\ $\ge$ 1 \AA\ at 1.0 $<z<$ 1.86. However, the \mgii\ 
line strength is generally found to be correlated with the reddening. For example, \citet{budzynski2011} have found that the dependence of \wmg\ on \ebv\ can be modeled by a 
power-law for 1 $\le$ \wmg\ (\AA) $\le$ 5, and $\sim$83\% of the high confidence 2175 \AA\ candidates from \citet{jiang2011} have \wmg\ $\ge$ 2 \AA\ and \wfe\ $\ge$ 1 \AA. Recently, 
\citet{ledoux2015} have shown that dust content of C\,{\sc i} absorbers at 1.5 $<z<$ 4.5 selected from SDSS is significant, with mean \ebv\ = 0.065 and 30\% of them showing the 
2175\,\AA\ extinction feature. Moreover, C\,{\sc i} absorbers provide ideal targets to search for tracers of translucent molecular gas like CO \citep[e.g.][]{noterdaeme2011,noterdaeme2016}. 
\citet{noterdaeme2008} have found that the presence of \h2\ molecular absorption in $z>$ 1.8 DLAs is closely related to the dust content and metallicity. \hi\ \21\ absorption 
can also provide an efficient way to reveal cold and dusty absorbers \citep[e.g.][]{srianand2008}. Indeed a possible anti-correlation between the \hi\ \21\ spin temperature 
and gas phase metallicity has been reported \citep[see][and references therein]{kanekar2014a}. Here we investigate the dust content of strong \feii\ absorbers and its connection 
with \hi\ \21\ absorption, first in our present sample S1 (Section~\ref{sec_dust1}) and then in the full sample S2 (Section~\ref{sec_dust2}). 
\subsection{Dust content in sample S1} 
\label{sec_dust1}
We estimate the quasar reddening, \ebv, by fitting the quasar spectral energy distribution (SED) using the SDSS composite quasar spectrum \citep{vandenberk2001}, 
reddened by the Milky Way (MW), Small Magellanic Cloud (SMC), Large Magellanic Cloud (LMC) or LMC2 supershell extinction curves \citep{gordon2003}. We follow the 
same procedure as detailed in the works of \citet{srianand2008} and \citet{noterdaeme2009a,noterdaeme2010b}. In case of multiple epoch spectra available of the same 
quasar, we fit the SED of all of them and take the fit which gives the minimum $\chi^2$. We find that the \ebv\ values obtained from fitting multiple epoch spectra 
of the same quasar can differ by $\sim$10$-$20\%. This should be taken as the typical systematic error in the \ebv\ values. We note that in the majority of cases, a
better fit was obtained for SDSS-DR7 spectra than for SDSS-Baryon Oscillation Spectroscopic Survey (BOSS) spectra of the same quasar. This could be due to flux 
calibration overestimate of the blue side of the BOSS spectra.

\begin{figure*}
\subfloat[]{\includegraphics[height=0.35\textheight, angle=90]{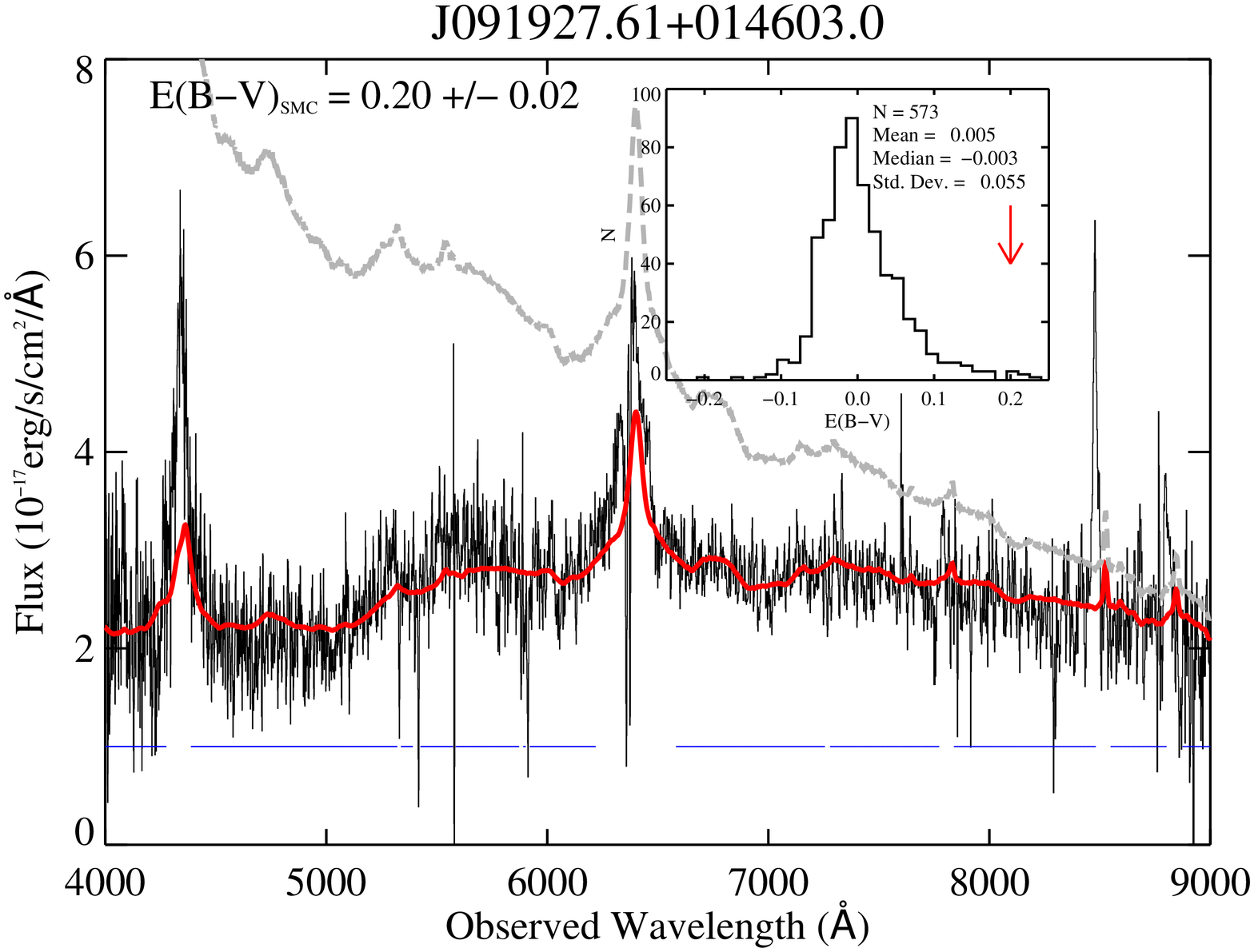} } 
\subfloat[]{\includegraphics[height=0.35\textheight, angle=90]{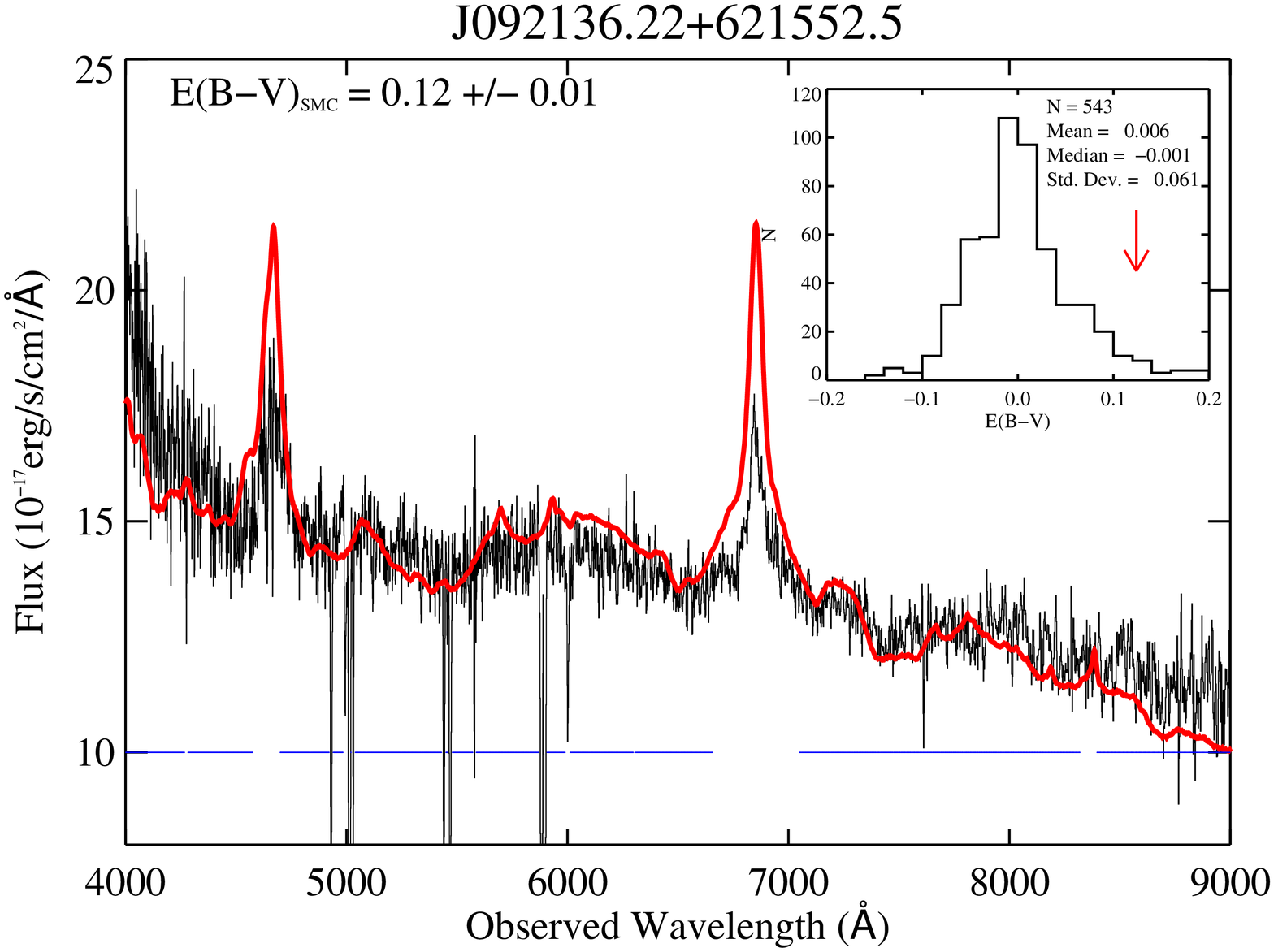} } \hspace{0.01cm}
\subfloat[]{\includegraphics[height=0.35\textheight, angle=90]{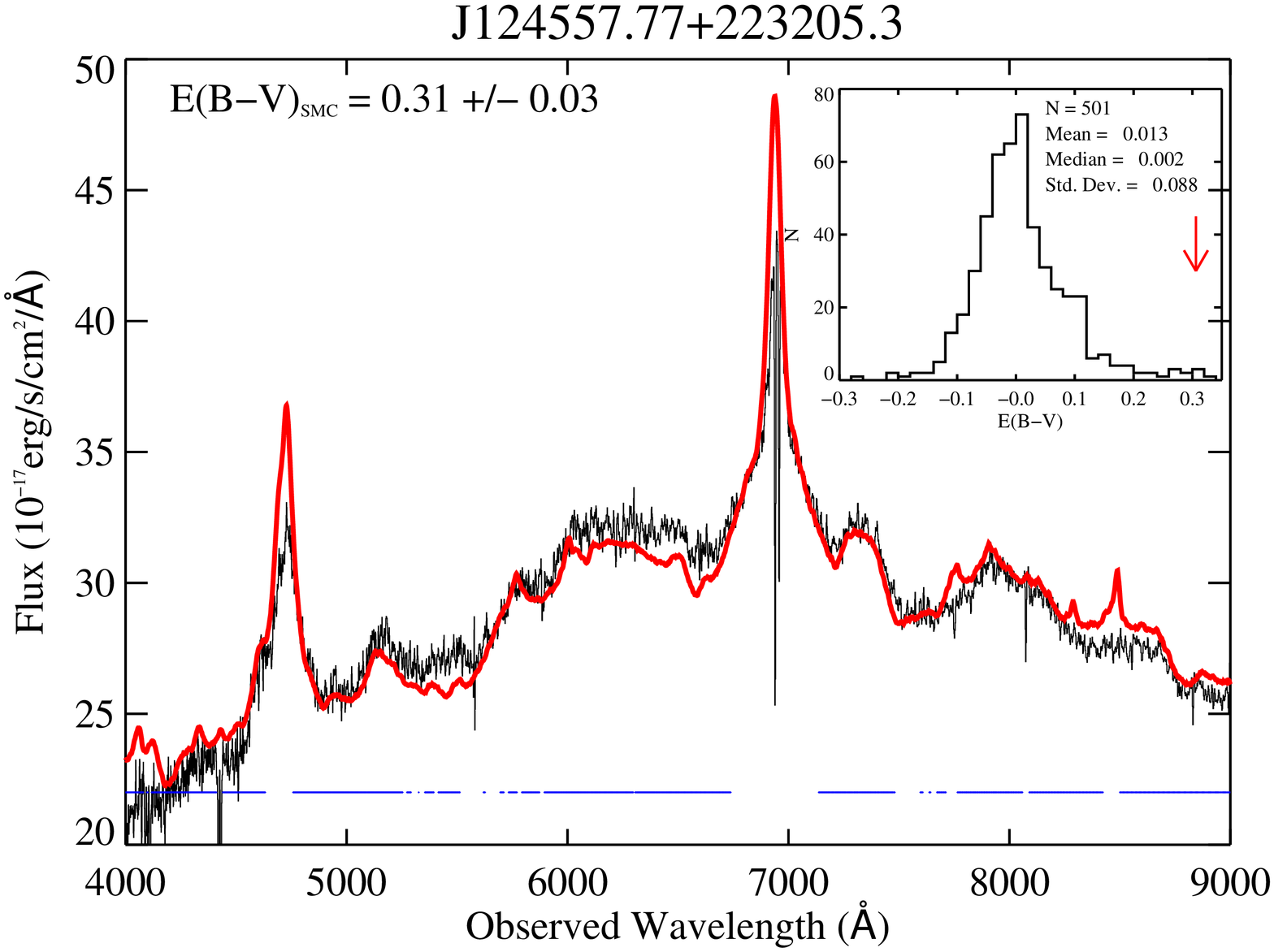} }
\subfloat[]{\includegraphics[height=0.35\textheight, angle=90]{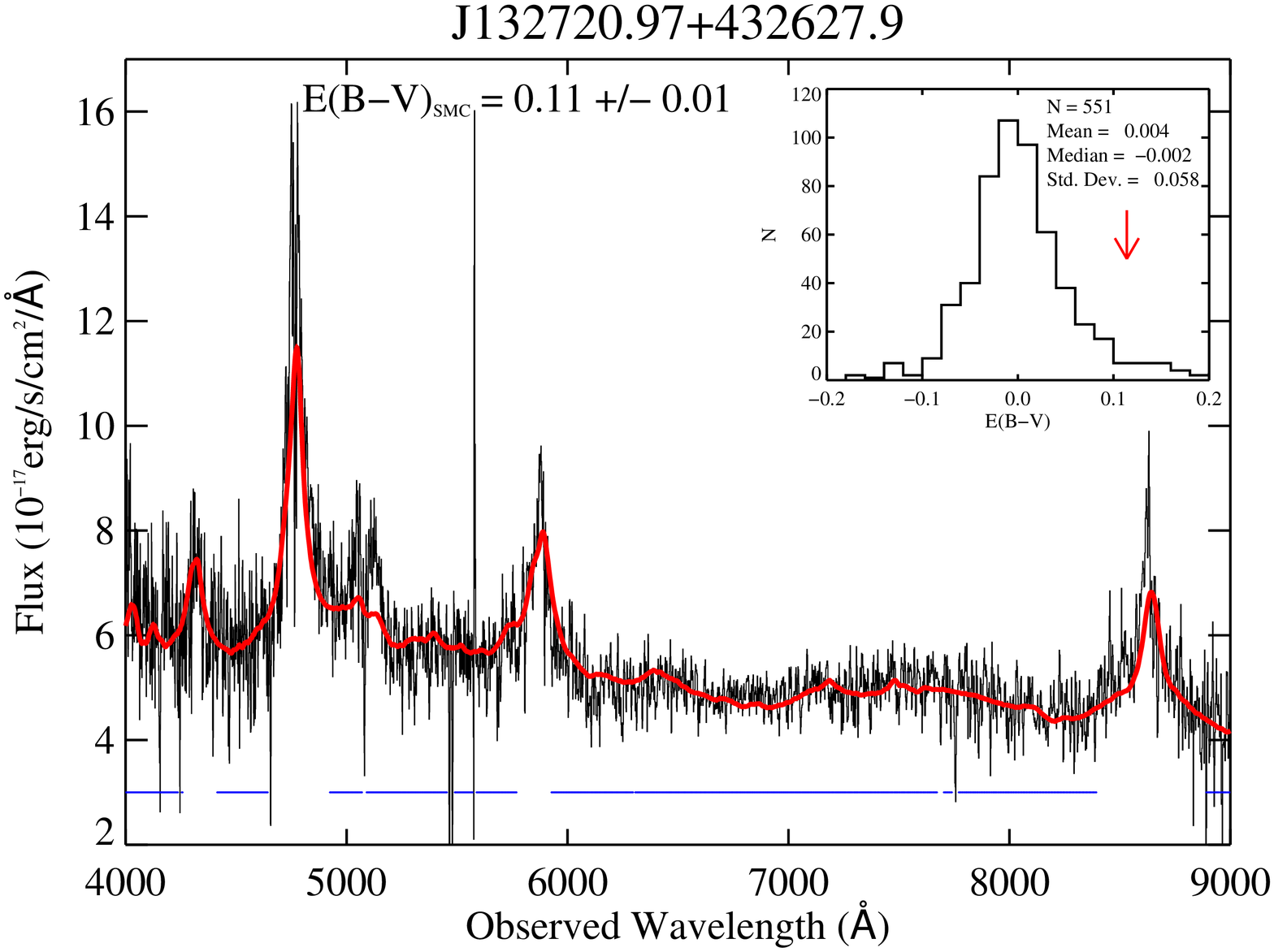} }
\caption{SED fit to the spectra of (a) J0919$+$0146, (b) J0921$+$6215, (c) J1245$+$2232, and (d) J1327$+$4326, taking into account the reddening by the intervening absorber. 
SMC extinction curve gives best fit to all of them. The best fit \ebv\ values are provided in each panel. The horizontal regions indicate the spectral range considered in the 
fitting process. Inset in each panel is the distribution of \ebv\ for a control sample of SDSS quasars (see Section~\ref{sec_dust1} for details). The number of quasars, mean, 
median and standard deviation of \ebv\ in the control sample are also provided. The arrows indicate the \ebv\ values obtained for the respective quasars. The unreddened SDSS 
quasar composite spectrum \citep{vandenberk2001} is shown {as the} gray dashed line in panel (a).}
\label{fig:sed}
\end{figure*}

We consider systems with \ebv\ $\ge$ 0.1 to be showing signatures of reddening, since below this the distribution of \ebv\ values could be dominated by the quasar 
SED-induced uncertainty. We find that 4 out of 16 quasars in our sample S1 (J0919$+$0146, J0921$+$6215, J1245$+$2232, and J1327$+$4326) have \ebv\ $\ge$ 0.1. \hi\ \21\ 
absorption has been detected towards three of these quasars (J0919$+$0146, J0921$+$6215 and J1327$+$4326). Note that the system towards J0919$+$0146 has been identified 
as a median-confidence (4.8$\sigma$) 2175 \AA\ absorber candidate by \citet{jiang2011}. The SED of these four quasars are best fit with SMC extinction law (see Fig.~\ref{fig:sed}). 
We note that the SED fit of J1245$+$2232 is not very good, which could be due to the complex intrinsic quasar emission. The errors provided for \ebv\ take into account 
the uncertainties in the extinction law parameters. For each of these four quasars, we applied the same SED fitting procedure with SMC extinction law to a control sample 
of SDSS non-BAL quasars within $\Delta z$ = $\pm$0.05$-$0.1 of \zem\ and $\Delta r_{mag}$ = $\pm$0.5$-$1.0 of $r_{mag}$ of the quasars and having spectra with signal-to-noise 
ratio $\ge$10. The details of the control samples are provided in Fig.~\ref{fig:sed}. The standard deviation of the \ebv\ values reflects the typical systematic error in 
the SED-fitting method due to the dispersion of the unreddened quasar SED. In case of the above four quasars, we find that their reddening is significant at $\sim2-4\sigma$ level. 

In Table~\ref{tab:dust}, we have estimated the \nhi\ for the above four systems in terms of the dust-to-gas-ratio relative to SMC ($\kappa$), using the observed mean 
relation between $A_V$ and \nhi\ in the SMC \citep{gordon2003}. Further, comparing this with the \taudv\ measurement we have estimated ($\kappa$ \ts)$/$\fc. In case of 
J0921$+$6215 and J1327$+$4326, where we have estimates of \fc\ (see Table~\ref{tab:radiosource}), we find that the \ts\ obtained for $\kappa$ = 1 is higher by a factor 
of $\gtrsim$ 2 than the upper limit on the gas kinetic temperature from the \hi\ \21\ line width (see Table~\ref{tab:gaussfit}). Hence, the extinction per hydrogen atom 
in these systems could be at least a factor of $\sim$ 2 higher than what is observed in the SMC. This could also be the reason for non-detection of \hi\ \21\ absorption 
towards J1245$+$2232. The possibility of the extinction per hydrogen atom being higher in high-$z$ absorbers has been suggested by G12. Further, from fig. 11 of \citet{ledoux2015}, 
it can be seen that in few of the high-$z$ \ci\ systems, the \ebv\ versus \nhi\ relation is consistent with the reddening per hydrogen atom being upto ten times higher 
than that seen in the Milky Way and the Magellanic clouds. This could imply a different grain chemistry or small grain size (i.e. larger total grain surface area) in 
these high-$z$ absorbers \citep[see e.g.][]{shaw2016,noterdaeme2016}. 
\begin{table}  
\centering
\caption{Parameters derived from SED fit to the quasars shown in Fig.~\ref{fig:sed} and discussed in Section~\ref{sec_dust1}}
\begin{tabular}{ccccc}
\hline
Quasar & \ebv\  & \nhi\                      & ($\kappa$ \ts)$/$\fc\ & $\kappa$ \\
       &        & $\times1/\kappa~10^{21}$ &                       &          \\
       &        & (\cms)                     & (K)                   &          \\
 (1)   & (2)    & (3)                        & (4)                   & (5)      \\
\hline
J0919$+$0146 & 0.20 $\pm$ 0.02 & 7  & 2500       & ---  \\
J0921$+$6215 & 0.12 $\pm$ 0.01 & 4  & 6000       & $>$3 \\
J1245$+$2232 & 0.31 $\pm$ 0.03 & 11 & $\ge$24000 & ---  \\
J1327$+$4326 & 0.11 $\pm$ 0.01 & 4  & 6000       & $>$2 \\
\hline
\end{tabular}
\begin{flushleft} {\it Notes.}
Column 1: quasar name. Column 2: \ebv\ obtained from SED fit. SMC extinction curve gives best fit in these cases. 
Column 3: \nhi\ in units of $1/\kappa~10^{21}$ (\cms) obtained from the relation between \nhi\ and $A_V$ in the SMC \citep{gordon2003}.
$\kappa$ is the dust-to-gas ratio relative to SMC.
Column 4: ($\kappa$ \ts)$/$\fc\ (K) obtained by comparing Column 3 with \taudv\ measurements given in Table~\ref{tab:radiopara}. 
Column 5: in case of J0921$+$6215 and J1327$+$4326 where we have estimates of \fc\ (see Table~\ref{tab:radiosource}), 
we constrain $\kappa$ by using the upper limit on $T_{\rm k}$ (see Table~\ref{tab:gaussfit}). 
\end{flushleft}
\label{tab:dust}
\end{table}
\subsection{Dust content in sample S2}
\label{sec_dust2}
\begin{figure*}
\subfloat[]{ \includegraphics[height=0.25\textheight, angle=90]{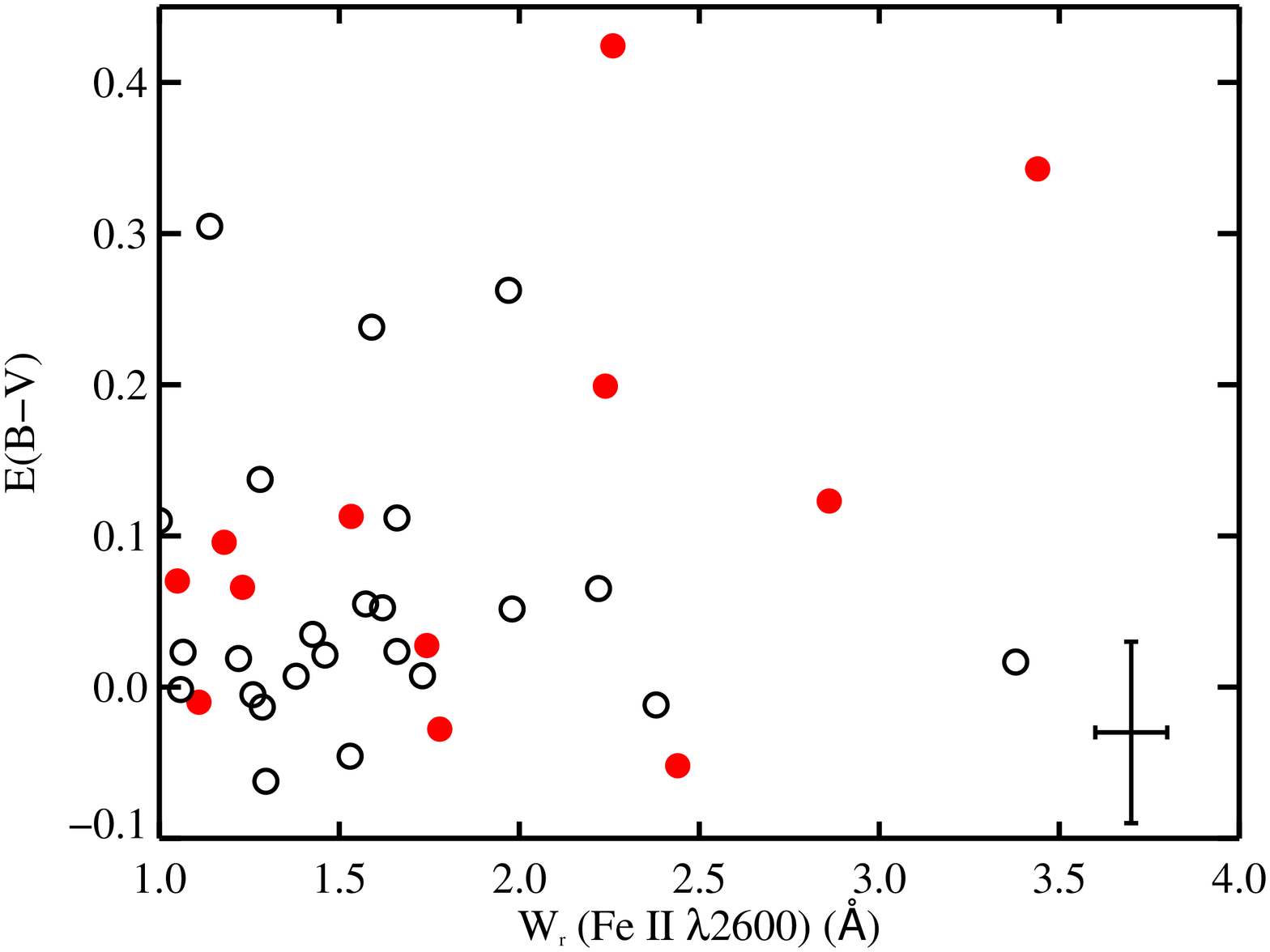} }
\subfloat[]{ \includegraphics[height=0.25\textheight, angle=90]{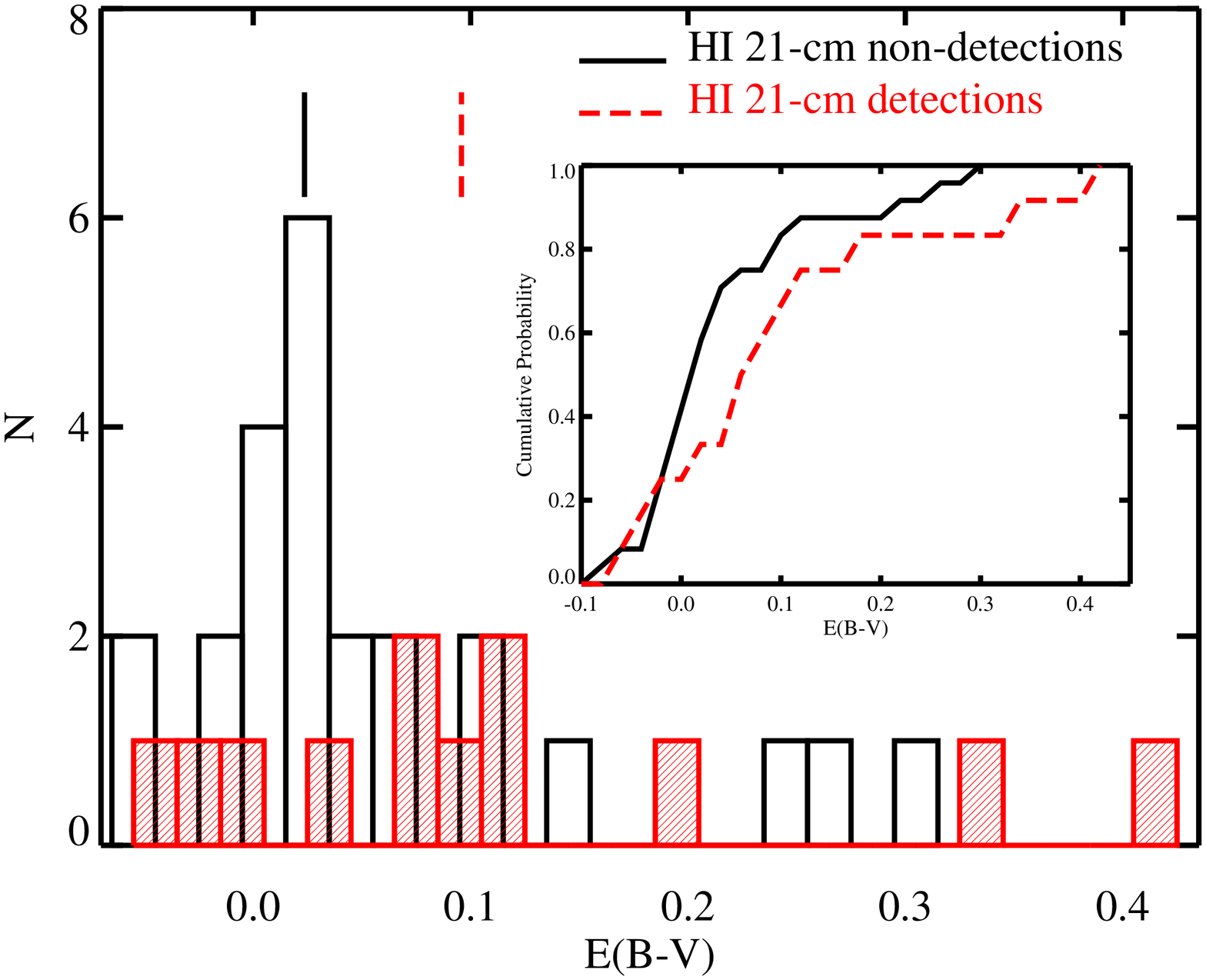} } 
\subfloat[]{ \includegraphics[height=0.26\textheight, angle=90]{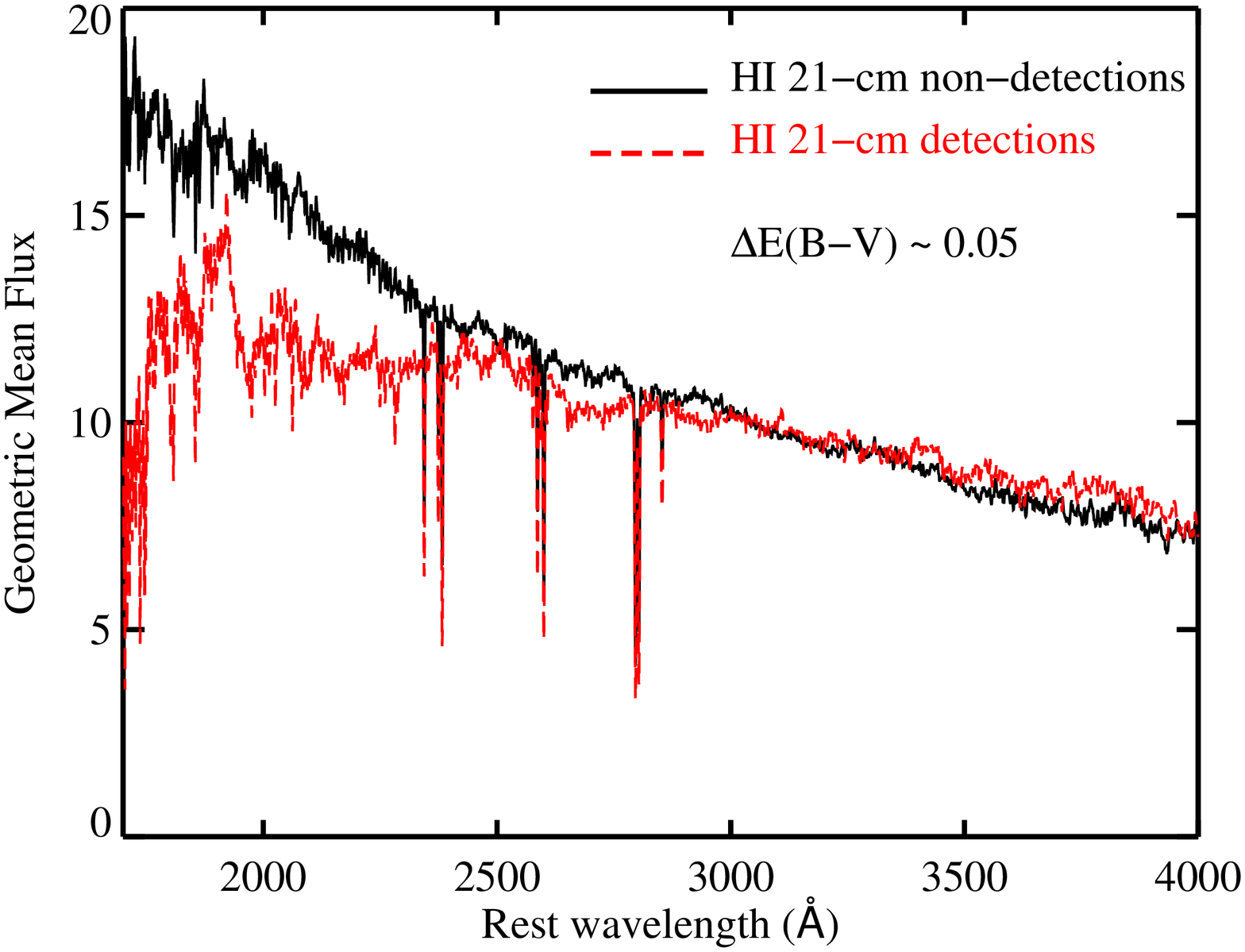} }
\caption{
(a) \ebv\ versus \wfe\ for the strong \feii\ systems. Filled symbols correspond to \hi\ \21\ detections and open symbols correspond to non-detections.
The typical error bars are plotted in the bottom right corner.
(b) The histogram distributions of \ebv\ values for strong \feii\ systems which show \hi\ \21\ absorption (red shaded) and those which do not (black open). 
The inset shows the cumulative distributions of \ebv\ for systems with and without \hi\ \21\ absorption as red dashed and black solid lines, respectively.
The vertical solid and dashed ticks mark the median \ebv\ for systems with and without \hi\ \21\ absorption, respectively.
(c) The geometric mean stacked spectrum of SDSS quasars with strong \feii\ systems at 0.5$<z<$1.5 (in rest frame) which are detected in \hi\ \21\ absorption is shown as the red dashed line, 
and the same for systems which are not detected in \hi\ \21\ absorption is shown as the black solid line. The spectra have been smoothed by 5 pixels for display purpose. 
The differential reddening, $\Delta$\ebv, is indicated in the figure (see Section~\ref{sec_dust2} for details).
}
\label{fig:ebv}
\end{figure*}

We estimated \ebv\ for all the strong \mgii\ systems that have been searched for \hi\ \21\ absorption (G09, G12). The \ebv\ of the strong \mgii\ systems is correlated with 
both \wmg\ and \wfe\ at 2$\sigma$ significance level. We find that strong \feii\ absorbers tend to cause more reddening, with median \ebv\ = 0.05 for systems with \wfe\ $\ge$ 
1 \AA, compared to median \ebv\ = 0.01 for systems with \wfe\ $<$ 1 \AA. A two-sided Kolmogorov-Smirnov (KS) test between the \ebv\ distribution of the strong and weak \feii\ 
systems suggests that the maximum deviation between the two cumulative distribution functions is $D_{\rm KS}$ = 0.50 with a probability of $P_{\rm KS}$ = 0.01 (where $P_{\rm KS}$ 
is the probability of finding this $D_{\rm KS}$ value or lower by chance). Hence, our strong \wfe\ based selection technique tends to select dusty absorbers. Though we note that 
among the strong \feii\ absorbers there is no correlation of \ebv\ with \wmg\ and \wfe\ (i.e. they are at $\lesssim1\sigma$ significance). We plot \ebv\ versus \wfe\ for the strong 
\feii\ systems in panel (a) of Fig.~\ref{fig:ebv}. In addition, we do not find any redshift evolution of \ebv.

Next, we check whether among the strong \feii\ systems, \hi\ \21\ absorption leads to more reddening in the quasar spectra. We show the histogram and cumulative distributions 
of the best-fit \ebv\ values for the systems with \hi\ \21\ detections and non-detections in panel (b) of Fig.~\ref{fig:ebv}. It can be seen that systems which show \hi\ \21\ 
absorption tend to cause more reddening in the quasar spectra. The median \ebv\ for \hi\ \21\ detections is 0.10, while it is 0.02 for the non-detections. A two-sided KS test 
between the two distributions gives $D_{\rm KS}$ = 0.42 with $P_{\rm KS}$ = 0.09. The connection between \hi\ \21\ absorption and dust is further illustrated in panel (c) of 
Fig.~\ref{fig:ebv}, which shows the geometric mean stacked SDSS quasar spectra of the \hi\ \21\ detections and non-detections. It can be clearly seen that \hi\ \21\ absorption 
on average causes more reddening in the quasar spectra. We fit the stacked spectrum of \hi\ \21\ detections considering the stacked spectrum of \hi\ \21\ non-detections as
a template. The best fit extinction curve is SMC, and the differential \ebv\ obtained is 0.05. In addition, we find that the \hi\ \21\ detection rate shows tentative evidence 
of increasing with reddening, i.e. \c21\ = 0.25$^{+0.17}_{-0.11}$ (0.25$^{+0.20}_{-0.12}$ for \t0\ = 0.2) for \ebv\ $<$ 0.1 and \c21\ = 0.38$^{+0.36}_{-0.20}$ (0.75$^{+0.73}_{-0.41}$ 
for \t0\ = 0.2) for \ebv\ $\ge$ 0.1. However, the increase of \c21\ with \ebv\ is not statistically significant due to the large uncertainties, and a larger sample is required 
to confirm this trend.    

Note that in the case of J0919$+$0146, unlike in the rest, we knew that the quasar showed signatures of reddening, based on the results of \citet{jiang2011}, before searching for \hi\ \21\ 
absorption. However, we find that the above results do not change beyond the statistical uncertainties on excluding this system, i.e. this system does not dominate the statistics. 
For example, on excluding this system, two-sided KS test between the \ebv\ distributions of \hi\ \21\ detections and non-detections gives $D_{\rm KS}$ = 0.39 with $P_{\rm KS}$ = 0.16, 
and the stacked spectrum of \hi\ \21\ detections show differential \ebv\ of 0.04 with respect to that of the non-detections. 

A correlation between \ebv\ and \nhi\ is expected based on observations in the Milky Way and the Magellanic clouds \citep{bohlin1978,gordon2003,gudennavar2012}.
However, we do not find any significant ($1.4\sigma$) correlation of \taudv\ with \ebv\ for the strong \feii\ systems. Finally, we note that 72\% of the 
systems with \ebv\ $\ge$ 0.1 have SMC type of dust. This is comparable to 74\% of the C\,{\sc i} absorbers having SMC type of dust \citep{ledoux2015}. Two out 
of the five \hi\ \21\ detections with \ebv\ $\ge$ 0.1 have LMC type of dust and are discussed in detail in \citet{srianand2008}. The remaining three 
have SMC type of dust and are discussed in Section~\ref{sec_dust1} (see also Fig.~\ref{fig:sed}). 
\section{Velocity width of \hi\ \21\ absorption}
\label{sec_velocity}
\subsection{Redshift evolution}
\label{sec_velocity1}
The 16 \hi\ \21\ absorption lines detected in the strong \feii\ systems (i.e. sample S2) exhibit a wide range of velocity widths, i.e. $v_{\rm 90}$ $\sim$8$-$160\,\kms (median = 56\,\kms), 
and FWHM of individual Gaussian components $\sim$3$-$130\,\kms (median = 23\,\kms). We compare the \hi\ \21\ absorption velocity width in the strong \feii\ systems with that in 
other intervening samples selected based on different criteria $-$ the sample of $z<$0.4 quasar-galaxy pairs \citep[QGPs; see][and references therein]{dutta2016b}, and samples 
of \mgii\ systems (which do not satisfy our criterion of \wfe\ $\ge$ 1 \AA\ at 0.5$<z<$1.5) (see G12 and references therein) and DLAs \citep[see][and references therein]{kanekar2014a,kanekar2014b}. 
Note that for this analysis we do not consider proximate absorbers with redshifts within $\sim$3000 \kms\ of the quasar redshift \citep{moller1998,ellison2002,prochaska2008b},
since the absorption could be associated with the quasar and intrinsic \hi\ \21\ absorption lines tend to be broader \citep{gereb2015,curran2016}. Hence, we do not consider the 
\hi\ \21\ absorption towards J0919$+$0146, which is $\sim$1700\,\kms\ blueshifted from the quasar emission redshift. This is the broadest \hi\ \21\ absorption line among all the 
\feii\ systems (the absorption can be fit with a single Gaussian component of FWHM = 131\,\kms). In addition, we do not consider the very broad (FWHM$\sim$235\,\kms) \hi\ \21\ 
absorption line detected towards J162439.09$+$234512.2 \citep{curran2007b}, as well as the \hi\ \21\ absorption lines detected towards gravitational lenses \citep{chengalur1999,kanekar2003}, 
where the velocity widths could be dominated by other factors like the radio structure.

Note that the different samples listed above probe different redshift ranges. When restricting to $z<$ 1, we do not find any difference in the $v_{\rm 90}$ distributions of the
galaxy-selected (or absorption-blind) sample of QGPs and the absorption-selected samples of \mgii$/$\feii$/$DLAs. Similarly, we do not find any difference in the $v_{\rm 90}$ 
distributions of the strong \feii\ systems and the other \mgii\ systems and DLAs, over 0.5$<z<$1.5. Note that the \taudv\ distributions in QGPs, DLAs, \mgii\ and \feii\ systems 
are also similar when we consider common redshift ranges. 

\begin{figure*}
\subfloat[]{ \includegraphics[height=0.35\textheight, angle=90]{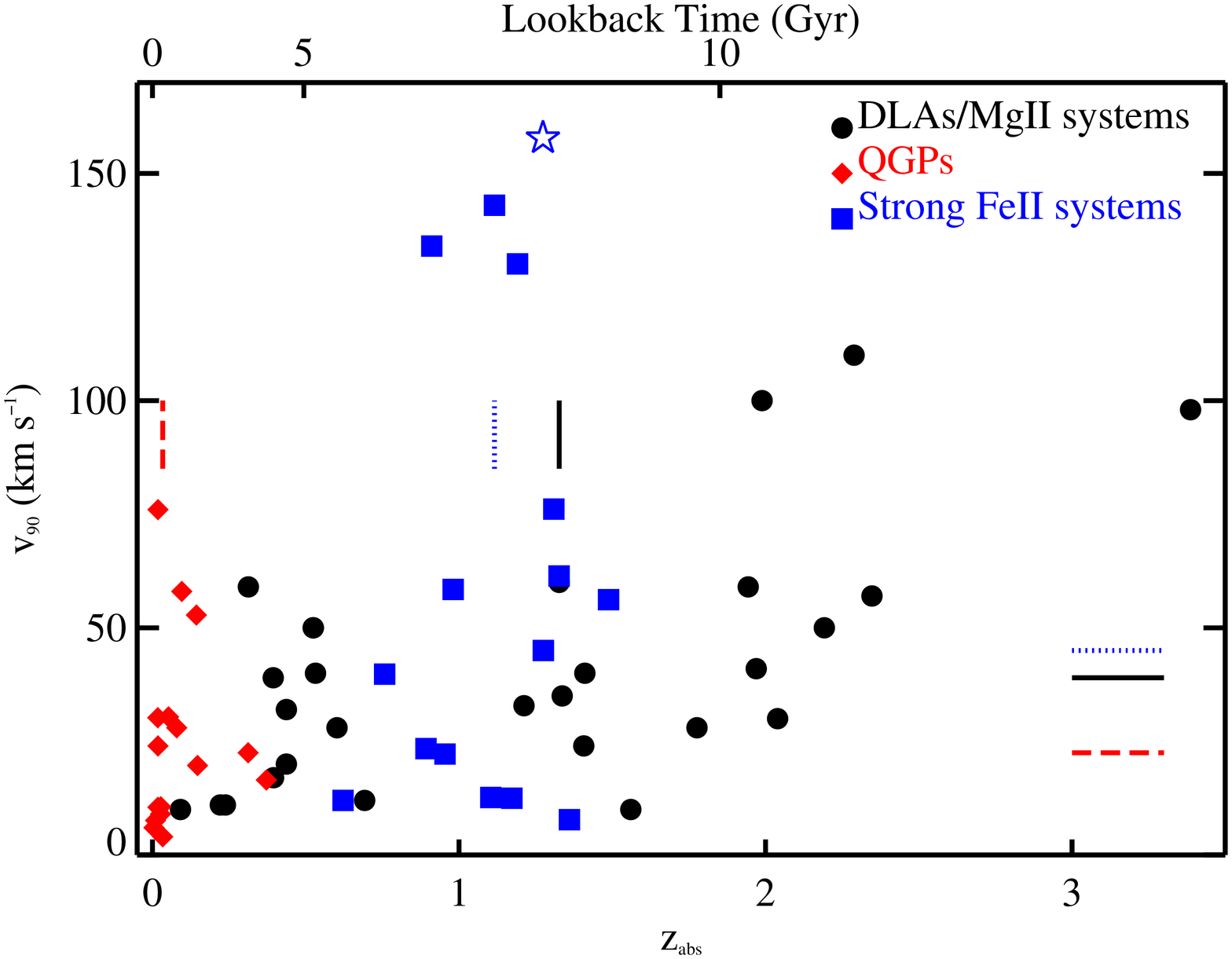} }
\subfloat[]{ \includegraphics[height=0.35\textheight, angle=90]{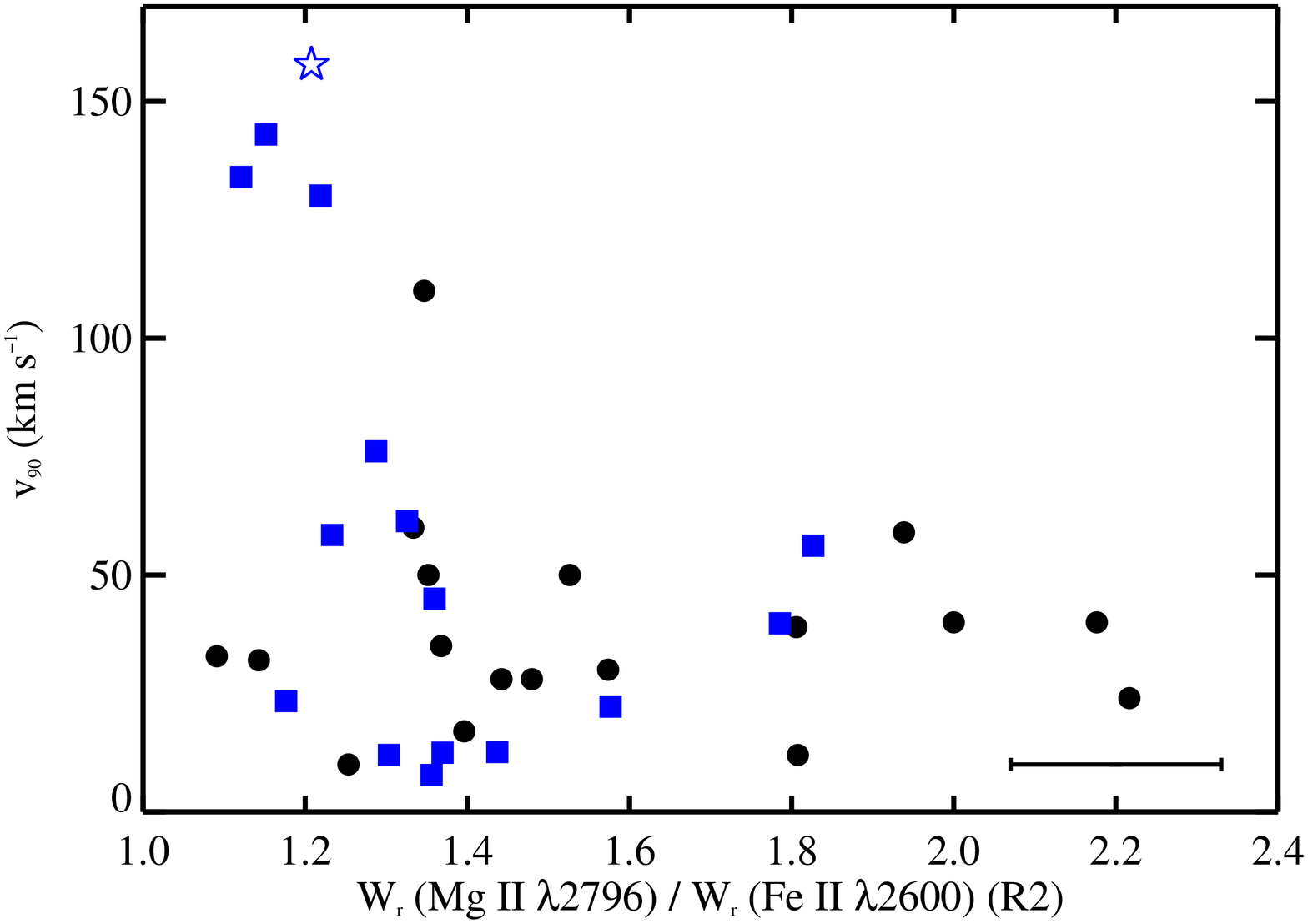} }
\caption{(a) The velocity width, $v_{\rm 90}$, of \hi\ \21\ absorption lines detected in samples of DLAs/\mgii\ systems (black circles), QGPs (red diamonds) and strong \feii\ systems (blue squares). 
The solid, dashed and dotted vertical (horizontal) ticks mark the median \zabs\ ($v_{\rm 90}$) for the DLAs/\mgii\ systems, QGPs and \feii\ systems, respectively.
The measurement towards J0919$+$0146 from our sample (which is not considered in the analysis; see text for details) is shown as a star.
(b) $v_{\rm 90}$ of \hi\ \21\ absorption lines detected at 0.3$<z<$2.3 as a function of the ratio R2. The symbol convention is same as in panel (a).
The typical error bar in R2 is plotted in the bottom right corner.
}
\label{fig:velocity}
\end{figure*}

The left panel of Fig.~\ref{fig:velocity} shows $v_{\rm 90}$ as a function of \zabs\ for the QGPs, the strong \feii\ systems and other \mgii\ systems and DLAs. It can be seen that 
$v_{\rm 90}$ of \hi\ \21\ absorbers shows an increasing trend with redshift. Test for correlation between $v_{\rm 90}$ and \zabs\ for all the \hi\ \21\ absorbers gives a positive 
correlation, with $r_{\rm k}$ = 0.34, $P(r_{\rm k})$ = 1.4 $\times$ 10$^{-4}$, $S(r_{\rm k})$ = 3.8$\sigma$ and $r_{\rm s}$ = 0.48, $P(r_{\rm s})$ = 1.4 $\times$ 10$^{-4}$, 
$S(r_{\rm s})$ = 3.6$\sigma$. Note that inclusion of the proximate systems and the systems from \citet{chengalur1999,kanekar2003,curran2007b}, leads to the significance of the 
anti-correlation becoming 3$\sigma$. On the other hand, if we consider only the measurements from absorption-selected samples (i.e. excluding the measurements from QGPs), the 
anti-correlation is significant at 2.5$\sigma$. 

The increasing trend of $v_{\rm 90}$ with redshift indicates that the velocity dispersion of \hi\ gas could be larger in high-$z$ galaxies. From fig. 4 of \citet{ledoux2006} it 
can be seen that for a given metallicity, the low-ionization metal lines in DLAs tend to be have larger velocity widths at $z>$2.4. The velocity-metallicity correlation in DLAs 
is believed to reflect the mass-metallicity relation in galaxies \citep{tremonti2004,moller2013,neeleman2013}. Moreover, fig. 3 of \citet{erb2006} shows that for a given metallicity, 
galaxies will have larger stellar masses at $z\sim$2 compared to at $z\sim$0.1. Hence, this may mean that for a given metallicity, the average velocity width is higher at high-$z$, 
as it originates from more massive galaxy haloes compared to at low-$z$. Therefore, one possible explanation of the $v_{\rm 90}$ vs. $z$ correlation is that a typical \hi\ \21\ absorber 
may be probed by larger mass haloes at high-$z$

In addition to $v_{\rm 90}$, the FWHM of individual Gaussian components of the \hi\ \21\ absorption profiles also show $4\sigma$ correlation with \zabs. The median FWHM (8\,\kms) of the 
\hi\ \21\ absorption lines detected at $z<0.75$ (which divides the sample into equal halves and corresponds to a lookback time of 6.6 Gyr) gives $T_{\rm k} \le$ 1400 K, while the median 
FWHM (18\,\kms) at $z\ge0.75$ gives $T_{\rm k} \le$ 7000 K. The large upper limits on the kinetic temperature indicates that non-thermal motions are most likely to dominate the line widths. 
The line widths could be driven by turbulent motion, which is closely linked with the supernova rate and hence the star formation rate \citep{maclow2004,elmegreen2004,mckee2007,tamburro2009}. 
Moreover, it is known that the star formation rate on average is higher at higher redshifts \citep{madau2014}. However, the \hi\ \21\ velocity width depends not just on the gas
structure and kinematics, but on the background radio source structure as well. To gain further insight into the redshift evolution of the \hi\ \21\ velocity width, we study its
dependence on various physical parameters next. 
\subsection{Dependence on physical parameters}
\label{sec_velocity2}
\citet{curran2007a} have reported a 2.2$-$2.8$\sigma$ level correlation between \hi\ \21\ absorption line widths and \wmg. However, G09 and K09 do not find similar correlation. 
Since \wmg\ and \wfe\ information cannot be uniformly obtained over the redshift range of \hi\ \21\ absorption measurements, we focus here on the systems at 0.3$<z<$2.3. 
For these systems, we do not find any correlation of $v_{\rm 90}$ of the \hi\ \21\ absorption lines with \wmg\ as well as \wfe\ (i.e. they are at $\lesssim1\sigma$ significance). 
The absence of correlation between the \hi\ \21\ absorption width and \wmg\ or \wfe\ can be explained if the \hi\ \21\ absorbing gas originates from only one or few of the 
metal line components, which are not necessarily the strongest. Indeed this has been observed in few cases where high resolution optical spectra are available \citep[G09;][]{srianand2012,rahmani2012,dutta2015}. 

Further, we do not find any correlation of $v_{\rm 90}$ with DR, R1 and R2 (i.e. they are at $\lesssim1\sigma$ significance). However, we find that the broadest \hi\ \21\ absorption 
lines occur when R2 is $\sim$1.2 (see right panel of Fig.~\ref{fig:velocity}). If we consider the measurement towards J0919$+$0146 (shown as a star in Fig.~\ref{fig:velocity}), 
there is a weak 1.5$\sigma$ anti-correlation between $v_{\rm 90}$ and R2 (significance reduces to 1.2$\sigma$ on excluding it). Assuming that both \mgii\ and \feii\ absorption trace 
the same velocity field and $b$ parameters in the range 40$-$60 \kms\ (as indicated by our analysis in Section~\ref{sec_metals2}), from single-cloud curve-of-growth we find that 
values of R2 $\lesssim$1.2 will occur in the saturated part where $N$(\feii) $>$ 3 $\times$ 10$^{15}$ \cms. We can put a lower limit on \nhi\ $>$ 10$^{20}$ \cms\ assuming solar 
metallicity and solar abundance of Fe \citep{asplund2009}. Since Fe is usually prone to depletion onto dust grains \citep{savage1996,draine2003,jenkins2009}, this is a strict 
lower limit on \nhi. Hence, such large $N$(\feii) are expected from self-shielded regions with high \nhi. The probability of encountering multiple cold \hi\ clouds along such 
sightlines would be high, which can explain the observed broad \hi\ \21\ lines. On the other hand, when R2 is higher, \nhi\ would be lower and in turn the probability of encountering 
multiple cold gas clouds along such sightlines would be lower. This can explain the low $v_{\rm 90}$ values when R2 $\ge$ 1.5. The scatter in the $v_{\rm 90}$ values when R2 is close 
to 1 can be explained if \hi\ \21\ absorption does not always arise from all the metal line components, as also discussed above. 

As noted in Section~\ref{sec_velocity1}, the velocity width of the \hi\ \21\ absorption lines also depends on the radio structure of the background sources. We note that G12 have 
not found any correlation of $v_{\rm 90}$ with the linear size of the radio sources measured from VLBA images. However, the three \feii\ systems with $v_{\rm 90}$ $>$ 100\,\kms\ are towards 
radio sources with extended morphology at arcsecond-scales (G12). Note that in case of the broad proximate absorber towards J0919$+$0146, the radio source is compact in our GMRT image. 
Even when the radio sources are compact at arcsecond-scales, if radio structures are present at parsec-scales, that could lead to broadening of the line \citep[e.g.][]{srianand2013,srianand2015,biggs2016}. 
Hence, we emphasize the need for multi-wavelength VLBA sub-arcsecond-scale images of all the radio sources towards which \hi\ \21\ absorption lines are detected in order to characterize 
their radio structure and address the origin of the velocity widths. 
%
%
\section{Summary}  
\label{sec_summary} 
Using strong \mgii\ systems (\wmg\ $\ge$ 1\,\AA) having measurements of both \nhi\ and \wmg\ at $z<1.65$ and $z\sim2$, we have found that the probability
of having log~\nhi\ $\ge$ 20.3, i.e. DLA column densities, can be increased by a factor of $\sim$1.4$-$1.7, by selecting strong \feii\ systems with \wfe\ 
$\ge$ 1\,\AA. Hence, we have searched for \hi\ \21\ absorption in a sample of strong \feii\ systems at $0.5<z<1.5$ selected from SDSS-DR12, using GMRT
and GBT. We have detected \hi\ \21\ absorption in six of these systems.

Combining our sample with that of strong \feii\ systems from G09 and G12, we have estimated the detection rate of \hi\ \21\ absorption in strong \feii\ 
systems to be 0.30$^{+0.12}_{-0.09}$ for \t0\ = 0.3\,\kms. We have found that the detection rate increases with \wfe, being four times higher in systems 
with \wfe\ $\ge$ 1\,\AA\ compared to in systems with \wfe\ $<$ 1\,\AA. The detection rate of \hi\ \21\ absorption in strong \feii\ systems remains constant 
within the uncertainties over $0.5<z<1.5$. For \ts\ = 500 K (typical of DLAs at this redshift range) and \fc\ = 1, all the \hi\ \21\ absorption in strong 
\feii\ systems would arise from DLAs. Hence, a \wfe\ based selection appears to be efficient in detecting high \nhi\ cold gas. From the detection rate of 
\hi\ \21\ absorption in strong \feii\ systems and that of DLAs in strong \feii\ systems, we estimate the detection rate of \hi\ \21\ absorption in DLAs to 
be 0.67$^{+0.31}_{-0.23}$ at $0.5<z<1.5$ for \t0\ = 0.3\,\kms. This is three times higher than that estimated in $2<z<3.5$ DLAs and may indicate towards 
an increasing filling factor of cold gas in DLAs with time.

We do not find any significant correlation of the \hi\ \21\ absorption strength with the metal line properties. However, the metal absorption are systematically 
stronger in the stacked SDSS spectrum of the systems which show \hi\ \21\ absorption than in that of the \hi\ \21\ non-detections. In addition, the \hi\ \21\ absorbers 
tend to cause more significant reddening in the spectrum of the background quasars and there is a tendency for the detection rate of \hi\ \21\ absorbers to be higher 
towards more reddened quasars. The stacked spectrum of quasars with \hi\ \21\ absorption detected towards them is more reddened than that of quasars without any 
\hi\ \21\ absorption. Hence, the above imply that \hi\ \21\ absorption is more likely to arise in metal rich dusty cold gas. Note that highly reddened systems are 
likely to be missed out in samples selected on the basis of optical/UV spectra. \citet{budzynski2011} have shown that \mgii\ samples, constructed from flux-limited 
quasar surveys, suffer 24\% and 34\% incompleteness for absorbers with \wmg\ $>$ 1 \AA\ and $>$ 2 \AA, respectively. Hence, upcoming blind \hi\ \21\ absorption surveys 
with the Square Kilometre Array pre-cursors could unravel a new population of dusty absorbers towards highly reddened quasars.

By comparing the velocity widths of intervening \hi\ \21\ absorption lines detected in samples of QGPs, DLAs and \mgii\ systems at $z<3.5$, we find evidence 
for the velocity widths to be increasing with redshift, which is significant at $3.8\sigma$. This could be because a typical \hi\ \21\ absorber may be originating
from a larger mass galaxy halo at high-$z$ compared to at low-$z$. However, the \hi\ \21\ velocity width depends on various factors like kinematics and structure 
of the absorbing gas as well as structure of the background radio source, and these need to be better understood in order to correctly interpret the redshift 
evolution. Finally, we emphasize the need for more \hi\ \21\ absorption detections to confirm different trends noted in this work with higher statistical significance. \\
%
%
\newline \newline

\noindent \textbf{ACKNOWLEDGEMENTS} \newline \newline
\noindent 
We thank the anonymous referee for useful comments.
We thank the staff at GMRT and GBT for their help during the observations. 
GMRT is run by the National Centre for Radio Astrophysics of the Tata Institute of Fundamental Research. 
GBT is run by the National Radio Astronomy Observatory (NRAO). The NRAO is a facility of the National Science Foundation operated under cooperative agreement by Associated Universities, Inc. 
NG, PN, RS and PPJ acknowledge the support from Indo-Fench centre for the promotion of Advanced Research (IFCPAR) under Project No. 5504-2.
JK acknowledges financial support from the Danish Council for Independent Research (EU-FP7 under the Marie-Curie grant agreement no. 600207) with reference DFF-MOBILEX--5051-00115.
This research has made use of the NASA/IPAC Extragalactic Database (NED) which is operated by the Jet Propulsion Laboratory, California Institute of Technology, 
under contract with the National Aeronautics and Space Administration. 

Funding for SDSS-III has been provided by the Alfred P. Sloan Foundation, the Participating Institutions, the National Science Foundation, and the U.S. Department of Energy Office of Science. 
The SDSS-III web site is http://www.sdss3.org/. SDSS-III is managed by the Astrophysical Research Consortium for the Participating Institutions of the SDSS-III Collaboration including the 
University of Arizona, the Brazilian Participation Group, Brookhaven National Laboratory, Carnegie Mellon University, University of Florida, the French Participation Group, the German Participation 
Group, Harvard University, the Instituto de Astrofisica de Canarias, the Michigan State/Notre Dame/JINA Participation Group, Johns Hopkins University, Lawrence Berkeley National Laboratory, Max Planck 
Institute for Astrophysics, Max Planck Institute for Extraterrestrial Physics, New Mexico State University, New York University, Ohio State University, Pennsylvania State University, University of 
Portsmouth, Princeton University, the Spanish Participation Group, University of Tokyo, University of Utah, Vanderbilt University, University of Virginia, University of Washington, and Yale University. 
%
%
\def\aj{AJ}%
\def\actaa{Acta Astron.}%
\def\araa{ARA\&A}%
\def\apj{ApJ}%
\def\apjl{ApJ}%
\def\apjs{ApJS}%
\def\ao{Appl.~Opt.}%
\def\apss{Ap\&SS}%
\def\aap{A\&A}%
\def\aapr{A\&A~Rev.}%
\def\aaps{A\&AS}%
\def\azh{A$Z$h}%
\def\baas{BAAS}%
\def\bac{Bull. astr. Inst. Czechosl.}%
\def\caa{Chinese Astron. Astrophys.}%
\def\cjaa{Chinese J. Astron. Astrophys.}%
\def\icarus{Icarus}%
\def\jcap{J. Cosmology Astropart. Phys.}%
\def\jrasc{JRASC}%
\def\mnras{MNRAS}%
\def\memras{MmRAS}%
\def\na{New A}%
\def\nar{New A Rev.}%
\def\pasa{PASA}%
\def\pra{Phys.~Rev.~A}%
\def\prb{Phys.~Rev.~B}%
\def\prc{Phys.~Rev.~C}%
\def\prd{Phys.~Rev.~D}%
\def\pre{Phys.~Rev.~E}%
\def\prl{Phys.~Rev.~Lett.}%
\def\pasp{PASP}%
\def\pasj{PASJ}%
\def\qjras{QJRAS}%
\def\rmxaa{Rev. Mexicana Astron. Astrofis.}%
\def\skytel{S\&T}%
\def\solphys{Sol.~Phys.}%
\def\sovast{Soviet~Ast.}%
\def\ssr{Space~Sci.~Rev.}%
\def\zap{$Z$Ap}%
\def\nat{Nature}%
\def\iaucirc{IAU~Circ.}%
\def\aplett{Astrophys.~Lett.}%
\def\apspr{Astrophys.~Space~Phys.~Res.}%
\def\bain{Bull.~Astron.~Inst.~Netherlands}%
\def\fcp{Fund.~Cosmic~Phys.}%
\def\gca{Geochim.~Cosmochim.~Acta}%
\def\grl{Geophys.~Res.~Lett.}%
\def\jcp{J.~Chem.~Phys.}%
\def\jgr{J.~Geophys.~Res.}%
\def\jqsrt{J.~Quant.~Spec.~Radiat.~Transf.}%
\def\memsai{Mem.~Soc.~Astron.~Italiana}%
\def\nphysa{Nucl.~Phys.~A}%
\def\physrep{Phys.~Rep.}%
\def\physscr{Phys.~Scr}%
\def\planss{Planet.~Space~Sci.}%
\def\procspie{Proc.~SPIE}%
\let\astap=\aap
\let\apjlett=\apjl
\let\apjsupp=\apjs
\let\applopt=\ao
\bibliographystyle{mnras}
\bibliography{mybib}
\end{document}